\documentclass[11pt]{article}

\usepackage[preprint]{acl}

\usepackage[utf8]{inputenc} 
\usepackage[T1]{fontenc}    
\usepackage{times}
\usepackage{latexsym}
\usepackage{inconsolata}
\usepackage{booktabs}       
\usepackage{amsfonts}       
\usepackage{nicefrac}       
\usepackage{microtype}      
\usepackage{xcolor}         
\usepackage{multirow}       
\usepackage{adjustbox}       
\usepackage{amsmath}
\usepackage{graphicx}
\usepackage{enumitem}
\usepackage{algorithm}
\usepackage{algpseudocode}
\usepackage{float}
\usepackage{wrapfig}
\usepackage[most]{tcolorbox}

\newtcolorbox{promptbox}[1][Prompt]{
  enhanced,
  breakable,
  colback=black!3,
  colframe=black!55,
  coltitle=white,
  fonttitle=\bfseries\small,
  fontupper=\small\ttfamily,
  boxrule=0.8pt,
  arc=2.5mm,
  left=2.5mm,
  right=2.5mm,
  top=3.5mm,
  bottom=2mm,
  before skip=0.75em,
  after skip=0.75em,
  title={#1},
  attach boxed title to top left={xshift=3.5mm,yshift*=-\tcboxedtitleheight/2},
  boxed title style={
    colback=black!55,
    colframe=black!55,
    boxrule=0pt,
    arc=1.5mm,
    left=2mm,
    right=2mm,
    top=0.8mm,
    bottom=0.8mm
  }
}

\title{CalBench: Evaluating Coordination--Privacy Trade-offs in Multi-Agent LLMs}

\author{
Chelsea Zou\thanks{Equal contribution.} \quad
Yiheng Yao\footnotemark[1] \quad
Selena She \quad
Noah D. Goodman \quad
Robert D. Hawkins \\
Stanford University \\
\texttt{\{cyzou, yaoyh, jshe, ngoodman, rdhawkins\}@stanford.edu}
}

\begin{document}

\maketitle

\begin{abstract}

Personal AI assistants are beginning to act as delegates with access to
calendars, inboxes, and user preferences. Calendar scheduling makes the trust
problem concrete: an assistant must coordinate with other assistants while
deciding what to reveal about the person it represents. We introduce CalBench, a
controlled benchmark for multi-agent calendar scheduling under private
information. In each task, $N$ agents manage separate private calendars and
schedule a stream of $M$ incoming meetings while minimizing disruption costs.
Because no agent can inspect another agent's calendar, success requires
language-mediated coordination rather than centralized planning. CalBench
generates solvable scenarios with CP-SAT oracle solutions and decentralized
non-LLM reference protocols, enabling evaluation of task success, excess cost,
communication efficiency, burden fairness, and privacy leakage under matched
information constraints. Across seven model families, we find that completion
alone misses important failures: agents leave avoidable cost on the table,
communication volume does not predict lower regret, and privacy-preserving
silence can deprive teammates of cost information needed for fair burden
allocation. CalBench provides a reproducible testbed for
studying whether autonomous assistants can coordinate on behalf of users before
deployment at scale. All traces and code can be found at \url{https://d3mern3a2mjjur.cloudfront.net/leaderboard} and \url{https://anonymous.4open.science/r/calbench2026-235F/README.md}.
\end{abstract}

\section{Introduction}

Calendar scheduling is a practical test of delegated agency. A scheduling
assistant represents a person with private constraints, preferences, and
commitments; when it negotiates with other assistants, it must act on that
person's behalf without exposing the whole calendar or making unreasonable changes.
This raises a new challenge: can assistants turn
private user state into shared commitments without leaking too much or imposing
avoidable costs on the people they represent?

Calendar scheduling is also \textit{structurally} multi-agent. Each participant
holds private information about availability, preferences, and commitments, and
no single party has full visibility or authority to impose a solution. Yet the
research community lacks rigorous benchmarks for \textit{necessarily}
multi-agent tasks with precise rewards. Many existing benchmarks do not cleanly
isolate decentralized coordination: even when a task involves multiple actors, a
centralized agent with full problem state can often solve it directly
\citep{cooperbench2026}. Calendar scheduling differs structurally. Each agent's
calendar is private and cannot be disclosed without violating the privacy
assumptions that make the task realistic; the multi-agent setup is the mechanism
by which agents selectively share enough information to coordinate while
withholding irrelevant private details.

\begin{figure*}[t]
    \centering
    \includegraphics[width=0.8\textwidth]{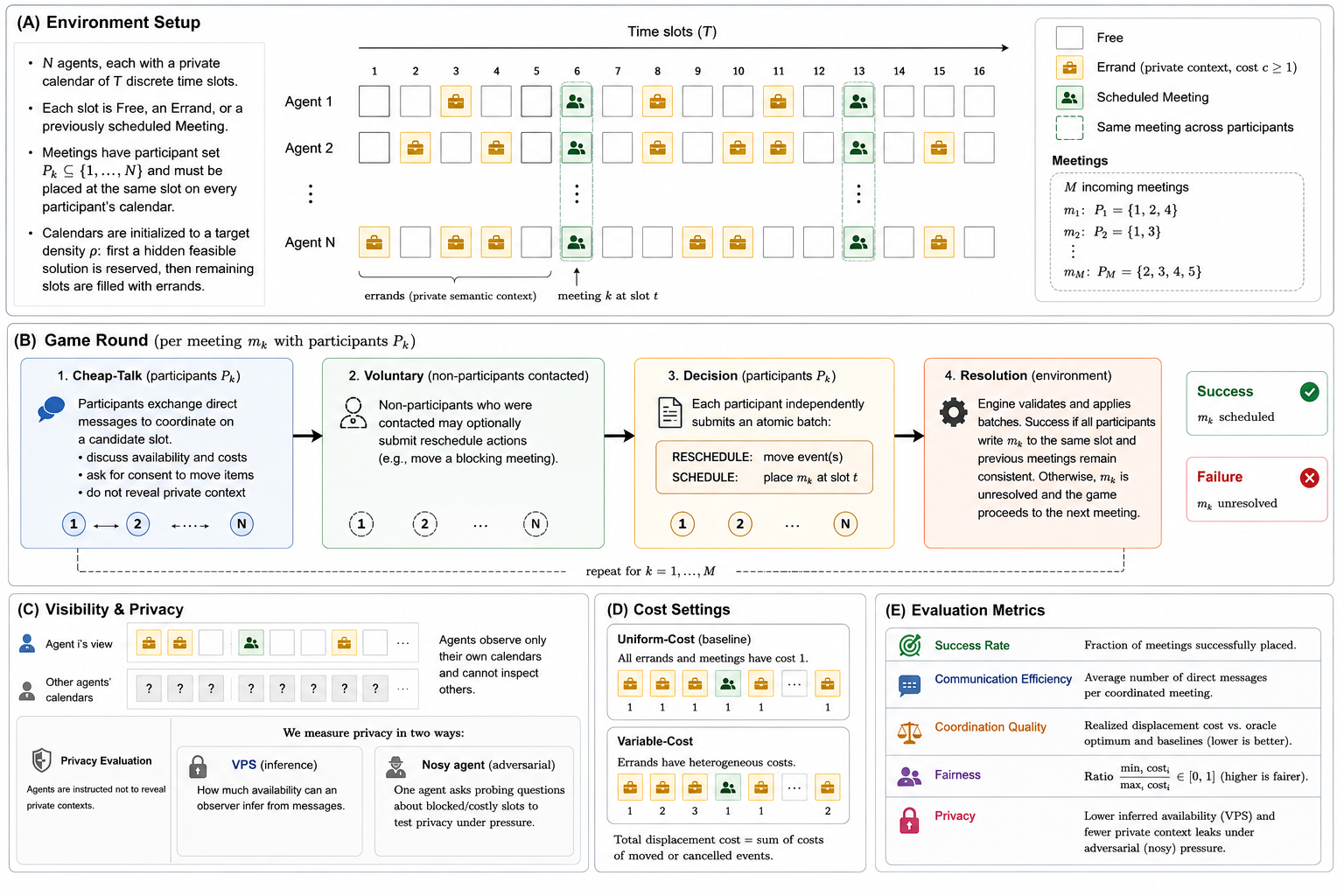}
    \caption{Overview of the CalBench environment. \textbf{(A) Environment setup:} Each of $N$ agents maintains a private calendar with $T$ discrete time slots containing free times, errands with private semantic contexts, and scheduled meetings. Meetings must be placed at the same slot across all participants' calendars. Calendars are initialized by reserving a hidden feasible solution and filling remaining slots to a global or per-agent target density.
    \textbf{(B) Game round:} Each meeting is processed through four phases: cheap-talk over private DMs, participant groupchat, or all-agent groupchat; voluntary rescheduling by contacted non-participants; independent decision submission via atomic action batches; and environment-level resolution and validation. 
    \textbf{(C) Visibility and privacy:} Agents observe only their own calendars. Privacy is evaluated through inference-based leakage (VPS), with semantic-context leakage audited separately in the appendix.
    \textbf{(D) Cost settings:} We study both uniform-cost and variable-cost environments, where errands may carry heterogeneous displacement costs.
    \textbf{(E) Evaluation metrics:} Runs are scored by scheduling success, excess displacement cost relative to the oracle, communication volume, burden fairness across agents, and privacy leakage.}
    \label{fig:overview}
\end{figure*}

We present \textbf{CalBench}, a benchmark for multi-agent calendar scheduling
that generates scenarios with oracle solutions via CP-SAT \citep{cpsatlp} and
non-LLM reference protocols under the same private-information constraints. Agents
coordinate through configurable cheap-talk channels \citep{crawford1982strategic}
and independently commit scheduling actions; the environment, phase structure,
and evaluation metrics are summarized in Figure~\ref{fig:overview}. The harness
supports three communication conditions: private agent-to-agent DMs,
meeting-participant groupchat, and all-agent groupchat. We evaluate seven models
on a mixed-agent benchmark suite.
Our contributions are: (1) a multi-agent calendar scheduling benchmark with
precise, computable evaluation criteria; (2) a scenario-generation harness that
produces controllable, solvable tasks with known oracle metrics and CP-SAT
feasible-fraction difficulty buckets; and (3) analysis of seven models covering
task success, coordination quality, communication efficiency, fairness, privacy
leakage, and recurrent language-level failure modes.

\section{Related Work}

\subsection{Agentic and multi-agent LLM benchmarks}

Recent benchmarks evaluate LLM agents in interactive settings, including tool
use and decision-making \citep{liu2024agentbench}, tool-agent-user interaction
\citep{yao2024taubench,barres2025tau2}, and natural-language planning with full
task context \citep{zheng2024naturalplan}. Multi-agent benchmarks study social
intelligence, collaboration, competition, and communication structure
\citep{zhou2024sotopia,zhu2025multiagentbench,mahmud2025collab}, with recent
work showing that multi-agent systems can underperform single-agent baselines
\citep{cooperbench2026,davidson2025collaborationgap,vink2025reproducibility}.
Partial-information work such as CRAFT further shows that communication skill
need not translate into collective success \citep{nath2026craft}. CalBench
extends this line with a verifiable scheduling benchmark combining private
information, oracle solutions, cost-sensitive objectives, fairness, and
trace-level communication analysis.

\subsection{Agents as user delegates}

Recent work frames assistant agents as delegates in a principal-agent
relationship: they must complete a task while exercising care, loyalty, and
confidentiality toward the user they represent. SocialReasoning-Bench applies
this framing to calendar coordination and marketplace negotiation, measuring
whether an agent secures value for its user and follows a reasonable decision
process \citep{microsoft2026socialreasoning}. CalBench studies a different
failure surface. Rather than a single assistant negotiating with a fixed
counterparty, CalBench places several assistants in a shared scheduling problem
where each agent has private state, no agent has global authority, and agreement
must remain consistent across all affected calendars. This adds group-level
metrics for feasibility, excess disruption, burden allocation, and privacy
leakage through the language used to coordinate.

\subsection{Negotiation and bargaining}

Negotiation under private utilities is closely related to CalBench. Prior work
studies neural negotiators with private rewards \citep{lewis2017deal},
strategic dialogue in Diplomacy \citep{fair2022diplomacy}, and LLM negotiation
benchmarks over bargaining, trading, and multi-issue games
\citep{bianchi2024negotiationarena,abdelnabi2024llmdeliberation}. Meeting
scheduling has also been framed as negotiation among autonomous agents
\citep{renting2024meeting}. CalBench differs by treating scheduling as
cooperative coordination over temporal constraints: agents must reach agreement,
minimize disruption relative to an oracle optimum, and maintain consistent
commitments.

\subsection{Distributed constraint optimization and calendar scheduling}

Calendar scheduling is a distributed constraint optimization problem, where
agents coordinate over private calendars, participant constraints, and global
utility \citep{maheswaran2004dimes,enembreck2012mulbs,modi2004multiagent}.
Prior work includes DiMES for distributed multi-event scheduling
\citep{maheswaran2004dimes}, MULBS for collaborative meeting scheduling
\citep{enembreck2012mulbs}, and rescheduling models where later high-priority
meetings may displace prior commitments \citep{modi2004multiagent}. CalBench
adopts this DCOP framing but uses a streaming regime: meetings arrive
sequentially against calendars already filled with earlier commitments. We
therefore compare LLM agents against decentralized non-LLM reference protocols
and use a full-information global oracle to define optimal cost
(\S\ref{sec:baselines}).

\subsection{Privacy in distributed coordination}

Privacy is central to distributed scheduling, yet negotiation can still leak
private information \citep{greenstadt2006privacy,brito2008privacy,farhadi2021faithful}.
We use Valuations of Possible States (VPS) \citep{maheswaran2006vps} as our
numeric-disclosure metric (\S\ref{sec:vps}). LLM-agent systems introduce further
privacy risks under contextual integrity \citep{nissenbaum2004ci}, including
leaks through assistants, action traces, tools, prompt injection, inter-agent
messages, shared memory, and tool arguments
\citep{mireshghallah2024confaide,shao2024privacylens,bagdasaryan2024airgap,debenedetti2024agentdojo,elyagoubi2026agentleak}.
CalBench measures both \emph{utility leakage} through availability and cost
disclosures, and \emph{semantic leakage} through unnecessary event details. This
targets multi-agent contextual privacy \citep{magpie2025}, not just output-only
disclosure.

\section{The CalBench Environment}
\label{sec:env}

Each CalBench game consists of $N$ agents, $T$ discrete time slots, and a stream
of $M$ incoming meetings. Agent $i$ maintains a private calendar whose slots may
be free, occupied by a movable errand, occupied by a previously scheduled
meeting, or blocked by an immovable commitment. Occupied entries carry a
displacement cost and a private semantic context visible only to the owning
agent. Each incoming meeting $k$ has a participant set
$P_k \subseteq \{1,\ldots,N\}$ and must be written to the same slot on every
participant's calendar, so agents must coordinate from local views and received
messages rather than from a shared global state.

\textbf{Calendar generation.}
The task generator first samples a hidden feasible witness assignment for the incoming meetings. It then fills calendars with errands while preserving enough free absorbing slots for the witness schedule to remain
feasible. Calendar \textit{density} controls the fraction of each calendar initially occupied by errands. Some calendar commitments are immovable (blocked) so agents must distinguish flexible events from hard constraints and avoid schedules that rely on impossible displacement. CalBench supports shared and per-agent density settings, allowing tasks in which some agents represent humans with substantially denser calendars than others. In our experiments, starting configurations are balanced across models for fair evaluation. 

\textbf{Costs.}
Previously scheduled meetings have displacement cost \(1\) for each participant calendar on which the meeting must be moved. In the uniform-cost setting, every errand also has displacement cost \(1\). In the varied-cost setting, errands are assigned low, medium, or high disruption costs from a balanced set. Internally, these correspond to \(\{1,2,3\}\), but agents are shown a logarithmic scale \(\{1,10,100\}\).\footnote{We use a logarithmic presentation because small linear differences such as \(\{1,2,3\}\) were not behaviorally salient in pilot runs: models often treated these costs as qualitatively similar. The displayed log scale makes the intended preference ordering clearer to agents, while we map costs back to the linear scale for leaderboard ranking.} The varied setting also creates cases where the best solution requires
moving previously scheduled meetings instead of expensive errands, so agents should reach out to external meeting participants from prior meetings. Thus, CalBench tests whether agents balance preferential costs, recognize when external coordination is worthwhile, and leak private information while doing so.

\textbf{Private contexts and sensitivity.}
Each errand and prior meeting is linked with a natural-language private context before it is shown to an LLM agent. Contexts are drawn from label banks with three sensitivity tiers: public, sensitive, and very sensitive. Agents see the private labels in their own calendar render, but they do not see the underlying tier labels. These contexts are not needed to solve the scheduling problem: agents may need to communicate availability, but not the private reason why a slot is occupied. This design lets the harness separate two forms of privacy loss. VPS measures inference about private calendar states (i.e., blocked slots and movement costs), while the context labels measure task-irrelevant semantic leakage in natural language. 

\textbf{Communication protocols.}
A game proceeds in rounds, one per incoming meeting. Each round begins with a
cheap-talk phase in which agents coordinate under one or more communication
channels. CalBench supports three message types:
\textsc{dm}, a private direct message to one recipient;
\textsc{participant-groupchat}, a shared channel visible only to the current
meeting participants; and \textsc{all-agent-groupchat}, a shared channel visible
to every agent in the task. Participants are active by default. Non-participants
become active only after receiving a direct message or observing an all-agent
groupchat message. This activation rule allows agents outside the current
meeting to help repair conflicts, while preventing unrelated agents from acting
unless the communication protocol has made them relevant.

\textbf{Round structure.}
After cheap-talk, the game enters a voluntary phase. In this phase, activated
non-participants may submit rescheduling actions, for example to move their copy
of a previously scheduled meeting that blocks the current participants. The
decision phase then asks each current participant to independently submit an
atomic batch of actions. A batch may contain \textsc{reschedule} actions followed
by a \textsc{schedule} action for the incoming meeting. Every reschedule action
must include a short justification explaining why the agent is moving its human
user's existing commitment. The batch is validated against the agent's calendar
and applied only if valid. Finally, the resolution phase checks that all meeting
participants scheduled the incoming meeting to the same slot and that any prior
meetings remain consistent across all of their participants' calendars. Failed
meetings are marked unresolved and the game proceeds to the next round.

\textbf{Task difficulty.}
Difficulty is based on CP-SAT feasibility score. For each task, we count how many distinct meeting-to-slot assignments satisfy the calendar constraints, and normalize by the total number of possible assignments. Lower scores correspond to harder tasks because fewer joint schedules are feasible. Because no two meetings may occupy the same slot, for $M$ meetings, $T$ total slots, and $F$ feasible assignments returned by CP-SAT, we define the feasibility difficulty score as
\[
d = \frac{F}{\frac{T!}{(T-M)!}}
\]

Tasks are then bucketed into easy, medium, and hard ranges based on this score.

\subsection{Non-LLM Reference Protocols}
\label{sec:baselines}

CalBench implements non-LLM reference protocols under the same private-information constraints. These protocols are not meant to imitate
open-ended LLM dialogue. They define interpretable operating points: full
numeric disclosure, binary feasibility exchange, and a tunable mechanism-design
frontier. \textsc{IMAP} sends full per-slot cost vectors to the initiator, who
picks the joint minimum: this is our low-cost, high-disclosure reference point.
\textsc{SD-MAP} \citep{modi2004multiagent} sends only typed proposal status per
slot and follows a scheduling-difficulty bumping rule: this is our low-disclosure
feasibility-first reference point. \textsc{DSM} \citep{farhadi2021faithful}
exposes a tunable privacy--cost curve through the interaction between offer-set
size and a per-offer privacy-cost weight $\theta$: lower privacy pressure permits
broader offer sets and leaks more score information, while higher privacy
pressure narrows offers, leaks less, and often pays higher coordination-adjusted
cost. Full specifications and inclusion rationales appear in
Appendix~\ref{app:baselines}.

\subsection{Privacy Leakage: VPS}
\label{sec:vps}

We measure structural privacy using Valuations of Possible States (VPS)
\citep{maheswaran2006vps}. For each meeting round $r$, target agent $i$, and
observer agent $j$, VPS measures how much the observer's belief about the
target's slot occupancy moves away from an uninformed prior $p_0=0.5$:
\begin{equation}
\mathrm{VPS}^{r}_{j \to i} \;=\; \sum_{k \in S} \bigl| \mathrm{Bel}^{r,\text{post}}_{j \to i}[k] - p_0 \bigr|.
\label{eq:vps}
\end{equation}
Because each slot contributes equally, VPS is reported in slot-equivalent units
and is comparable across uniform- and varied-cost calendars. We estimate VPS differently for typed protocols and LLM agents. For non-LLM
reference protocols, typed JSON messages have known meanings, so DSM score
vectors, IMAP cost vectors, and SD-MAP proposal/reply messages are converted
directly into slot-level evidence. For LLM agents, whose messages are
unstructured, we use reflection-calibrated VPS: after each round, a
measurement-only prompt asks each agent how its beliefs changed about every other
agent's calendar, and we count only belief movements that point toward the
target's true occupancy. This makes the LLM estimate conservative while keeping
it in the same slot-equivalent units as the typed-protocol estimator. We validate this proxy with an external LLM-as-judge audit over delivered chat
messages (Appendix~\ref{app:llm-judge-privacy-audit}). The judge finds higher
overall leakage but meaningful agreement with self-reflection, so we interpret
reflection-calibrated VPS as a lower bound on chat-observable privacy leakage.
VPS captures structural leakage about availability and costs; semantic-context
leakage, such as revealing private event descriptions, is audited separately.

\begin{figure*}[t]
    \centering
    \includegraphics[width=0.8\linewidth]{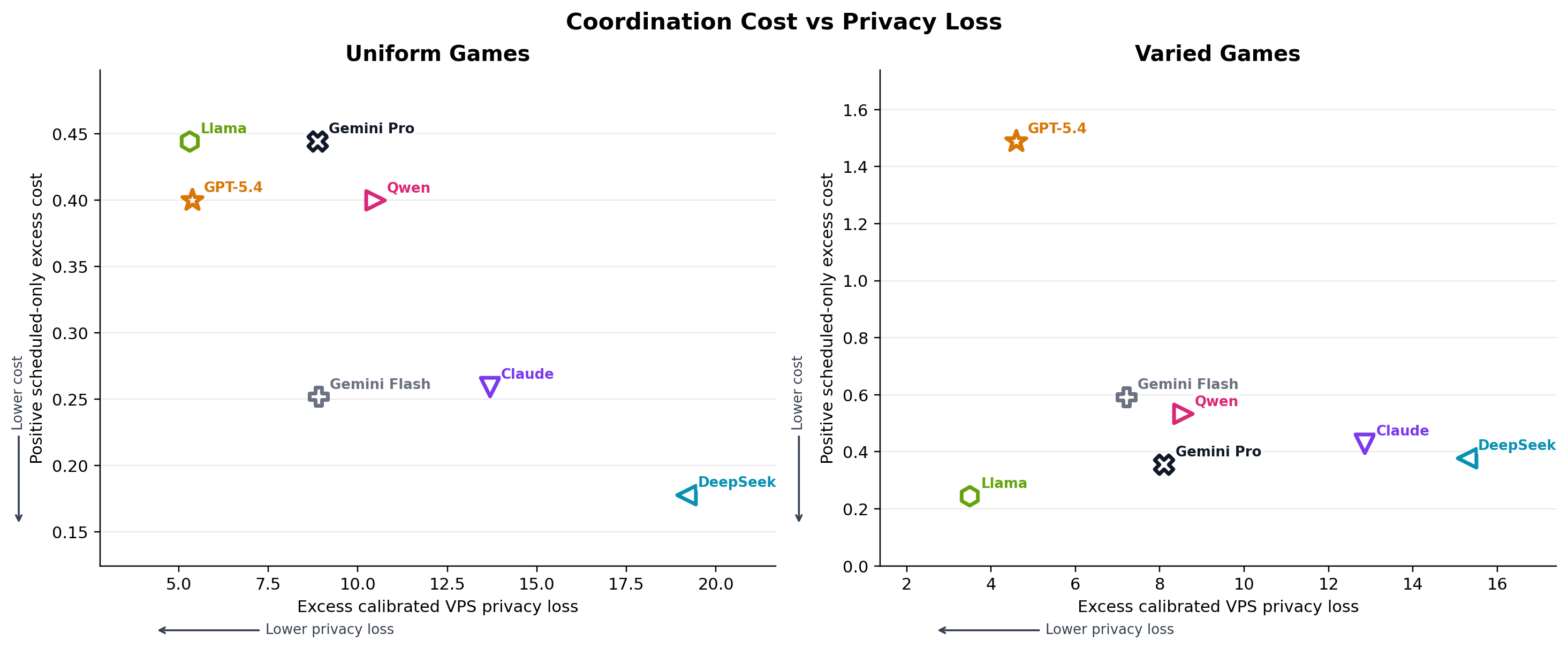}
    \caption{Privacy-cost plane for the canonical CalBench games, with uniform-cost games on the left and varied-cost games on the right. The closer to the origin, the better at reducing costs}
    \label{fig:baseline-frontier}
\end{figure*}

\section{Results}
\label{sec:results}

\renewcommand{\arraystretch}{0.8}

\begin{table*}[ht]
\centering
\scriptsize
\setlength{\tabcolsep}{4pt}
\resizebox{0.92\textwidth}{!}{%
\begin{tabular}{llrrrrrrr}
\toprule
Setting & Model & $N$ & Meetings & Coord. & Excess & Messages/mtg & Fairness & Excess cal. VPS \\
\midrule
Uniform & Claude Sonnet 4.6 & 135 & 2.54 & 84.69\% & 0.26 & 3.46 & 0.585 & 13.68 \\
 & Gemini 3.1 Pro & 135 & 2.58 & 85.93\% & 0.44 & 3.07 & 0.616 & 8.87 \\
 & Gemini 3 Flash & 135 & 2.59 & 86.17\% & 0.25 & 3.36 & 0.474 & 8.90 \\
 & GPT-5.4 Mini & 135 & 2.50 & 83.21\% & 0.40 & 3.08 & 0.541 & 5.38 \\
 & Llama 4 Maverick & 45 & 2.31 & 77.04\% & 0.44 & 3.11 & 0.529 & \textbf{5.30} \\
 & Qwen 3.6 Plus & 45 & \textbf{2.80} & \textbf{93.33\%} & 0.40 & \textbf{2.96} & 0.489 & 10.49 \\
 & DeepSeek V4 Pro & 45 & 2.69 & 89.63\% & \textbf{0.18} & 3.21 & \textbf{0.396} & 19.18 \\
\midrule
Varied & Claude Sonnet 4.6 & 135 & 2.53 & 84.20\% & 0.43 & 3.50 & 0.913 & 12.85 \\
 & Gemini 3.1 Pro & 135 & 2.45 & 81.73\% & 0.36 & 3.40 & 0.748 & 8.10 \\
 & Gemini 3 Flash & 135 & 2.36 & 78.52\% & 0.59 & 3.56 & 0.721 & 7.21 \\
 & GPT-5.4 Mini & 135 & 2.30 & 76.79\% & 1.49 & 3.44 & 1.376 & 4.60 \\
 & Llama 4 Maverick & 45 & 2.31 & 77.04\% & \textbf{0.24} & \textbf{3.20} & \textbf{0.627} & \textbf{3.49} \\
 & Qwen 3.6 Plus & 45 & 2.38 & 79.26\% & 0.53 & 3.76 & 0.662 & 8.56 \\
 & DeepSeek V4 Pro & 45 & \textbf{2.53} & \textbf{84.44\%} & 0.38 & 3.24 & 0.964 & 15.27 \\
\bottomrule
\end{tabular}%
}
\caption{Results on the canonical mixed-agent CalBench task set, split by cost setting. \(N\) is the number of evaluated agent-game seats. \emph{Meetings} is the mean number of successfully scheduled meetings per agent-game (max 3). \emph{Messages/mtg} counts direct, groupchat, and broadcast messages per successful participant-meeting. \emph{Excess} is positive scheduled-only excess displacement cost, \(\max(0, c^{\mathrm{real}}_{g,i} - c^{\mathrm{oracle}}_{g,i})\), using the 1/2/3 varied-cost scale. \emph{Fairness} is the absolute deviation of each agent's signed excess burden from the within-game mean. \emph{Excess cal. VPS} is calibrated slot-equivalent VPS. Lower is better for excess, messages, fairness, and VPS.}
\label{tab:main-results-combined}
\end{table*}

\subsection{Experimental Setup}
We evaluate a 90-task mixed-agent suite with 5 sequential meetings, 16 calendar
slots, \(N{=}5\) agents, and 3 rotating participants per meeting. The suite
contains 45 uniform-cost and 45 varied-cost tasks, with per-agent calendar
density sampled from \(\{0.6,0.8,1.0\}\) and blocked-errand counts from
\(\{2,4,6\}\). Table~\ref{tab:main-results-combined} evaluates four shared
models over three calibration slices (\(N=135\) seats per setting) and
Llama 4 Maverick, Qwen 3.6 Plus, and DeepSeek V4 Pro over their entrant slices
(\(N=45\) seats per setting). Fixed calibration runs use the same canonical task
suite; live-matchmaking and public leaderboard ratings use the OpenSkill
procedure in Appendix~\ref{app:openskill-ranking}.

\subsection{Metrics} We report five primary evaluation metrics for each model, shown in
Table~\ref{tab:main-results-combined}: task success, coordination cost, privacy cost, communication efficiency, and fairness (burden allocation). For all metrics, let \(\mathcal{G}_i\) be the set of games in which agent or
model \(i\) appears.

\emph{\textbf{Task success}} measures whether agents complete the requested scheduling
task. We define it as:
\[
S_i = \frac{1}{|\mathcal{G}_i|}
\sum_{g \in \mathcal{G}_i}
\frac{M_{g,i}^{\mathrm{sched}}}{M_{g,i}}
\]
where \(M_{g,i}\) is the number of meetings involving agent \(i\) in game \(g\)
and \(M_{g,i}^{\mathrm{sched}}\) is the number of those meetings successfully scheduled. A
meeting is successful only if all required participants write the meeting to the same slot and all previously scheduled meetings remain calendar-consistent.
This is the basic feasibility metric: an agent team that cannot consistently
place meetings has failed the core task, regardless of cost or privacy behavior.

\emph{\textbf{Coordination cost}} (excess cost) measures how much unnecessary displacement cost the
agents incur relative to an oracle scheduler. Let \(c_{g,i}^{\mathrm{real}}\) be the
realized displacement cost incurred by agent \(i\) in game \(g\), and let
\(c_{g,i}^{\mathrm{oracle}}\) be the CP-SAT oracle cost assigned to the same
agent for the same set of successfully scheduled meetings. We define scheduled
excess cost as
\[
C_i = \frac{1}{|\mathcal{G}_i|}
\sum_{g \in \mathcal{G}_i}
\max\left(0, c_{g,i}^{\mathrm{real}} - c_{g,i}^{\mathrm{oracle}}\right)
\]
When a game is only partially successful, we compute the oracle cost only over
the subset of meetings that were actually scheduled.

For Figure~\ref{fig:baseline-frontier}, however, scheduled-only excess cost can make failed coordination look artificially cheap: an unscheduled meeting creates
no displacement cost (\(c_{g,i}^{\mathrm{real}} = 0\)), even though it is a failure of the scheduling task. We
therefore plot a coordination-adjusted excess cost that penalizes unscheduled
meetings. For each task, we solve a second CP-SAT problem that keeps the same
feasibility constraints as the oracle but maximizes total displacement cost over
complete schedules. Let \(w_{g,i}\) be agent \(i\)'s cost in this worst complete
feasible schedule, and let \(o_{g,i}\) be the same agent's cost in the minimum-cost
complete oracle schedule. Let \(m_{g,i}^{\mathrm{miss}}\) be the count of
unscheduled participant-meetings for agent \(i\) in game \(g\). The plotted cost
uses an upper-bound regret proxy:
\[
\begin{aligned}
C_i^{\mathrm{adj}}
&= \frac{1}{|\mathcal{G}_i|}
\sum_{g \in \mathcal{G}_i}
\frac{\Delta c_{g,i}
+ m_{g,i}^{\mathrm{miss}}\Delta w_{g,i}}{M_{g,i}}, \\
\Delta c_{g,i}
&= \max\!\left(0, c_{g,i}^{\mathrm{real}} - c_{g,i}^{\mathrm{oracle}}\right), \\
\Delta w_{g,i}
&= \max\!\left(0, w_{g,i} - o_{g,i}\right).
\end{aligned}
\]
This keeps the unit as mean excess cost per meeting while assigning each missed meeting
an equal share of the task-level worst-complete regret proxy rather than dropping it from
the average. Because the penalty is the gap between the worst and best complete
feasible schedules for the whole task, it should be read as a conservative
coordination-failure penalty rather than the exact displacement cost of a
particular missed meeting.

\emph{\textbf{Privacy cost}} measures excess calibrated VPS privacy loss as
defined in Section~\ref{sec:vps}. VPS counts slot-equivalent belief movement,
giving all private slots equal weight so that model-to-model comparisons do not
depend on the task's displacement-cost scale. Lower VPS cost indicates better
privacy preservation.
\[
V_i = \frac{1}{|\mathcal{G}_i|}
\sum_{g \in \mathcal{G}_i} V_{g,i}
\]

\emph{\textbf{Communication efficiency}} measures how many messages are
exchanged to reach an outcome. For \(N_{g,i}^{\mathrm{msg}}\) communication
messages sent by agent \(i\), including direct messages, participant
groupchats, and all-agent broadcasts, we report messages per scheduled meeting:
\[
E_i = \frac{1}{|\mathcal{G}_i|}
\sum_{g \in \mathcal{G}_i}
\frac{N_{g,i}^{\mathrm{msg}}}{\max(M_{g,i}^{\mathrm{sched}},1)}
\]
Lower values indicate that agents reached outcomes with less communication
overhead and fewer opportunities for privacy leakage.

\emph{\textbf{Fairness}} measures whether excess displacement burden is evenly
distributed across agents within a game. We first compute each agent's excess
burden:
\[
e_{g,i} = c_{g,i}^{\mathrm{real}} - c_{g,i}^{\mathrm{oracle}}
\]
We then compute the mean excess burden within the game:
\[
\bar{e}_g = \frac{1}{N_g}\sum_{j=1}^{N_g} e_{g,j}
\]
Each agent's relative excess burden is
\[
r_{g,i} = e_{g,i} - \bar{e}_g
\]
We define the fairness cost for agent \(i\) as the average absolute relative
excess burden across games:
\[
F_i =
\frac{1}{|\mathcal{G}_i|}
\sum_{g \in \mathcal{G}_i} |r_{g,i}|
\]
Lower \(F_i\) indicates that the agent's excess burden is usually close to the within-game average, while higher \(F_i\) indicates that the agent often absorbs substantially more or less excess burden than its teammates.

\paragraph{Canonical mixed-agent tasks.}
Table~\ref{tab:main-results-combined} reports model results on the canonical
mixed-agent task set. In uniform-cost games, Qwen 3.6 Plus schedules the most
meetings (2.80 of 3; 93.33\%), while DeepSeek V4 Pro has the lowest excess cost
(0.18) and fairness cost (0.396). Uniform-cost excess costs are small overall,
ranging from 0.18 to 0.44.

The varied-cost setting separates models more clearly. DeepSeek V4 Pro has the
highest task success (84.44\%), but Llama 4 Maverick has the lowest excess cost
(0.24), fewest messages per successful participant-meeting (3.20), lowest
fairness cost (0.627), and lowest calibrated VPS (3.49). GPT-5.4 Mini performs
worst on varied-cost excess cost (1.49) and fairness (1.376), despite having low
VPS (4.60). Thus, completion alone does not capture schedule quality: models can
finish similar numbers of meetings while differing substantially in cost,
fairness, and privacy leakage. No semantic-context leakage was observed. For comparison, baseline protocol results are reported in Table \ref{tab:baseline-results}.

\begin{figure}[ht]
\centering
\includegraphics[width=\linewidth]{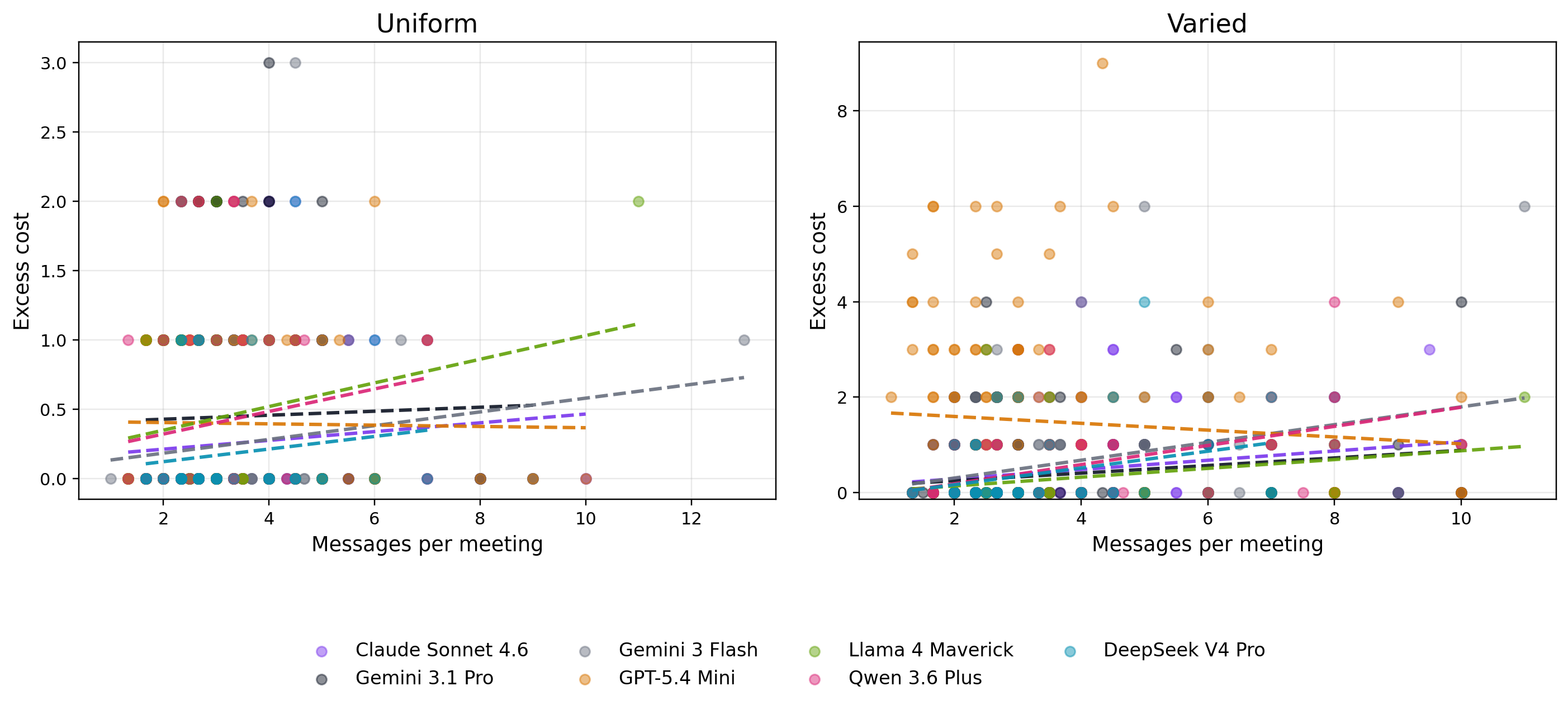}
\caption{Communication efficiency for the canonical model cohort. Each point is
one model seat in one game. The \(x\)-axis reports messages per successful
participant-meeting, and the \(y\)-axis reports positive scheduled-only excess
displacement cost using the 1/2/3 varied-cost scale. Dashed lines show
per-model OLS fits.}
\label{fig:communication-efficiency}
\end{figure}

\subsection{Message Volume Is Not Coordination Quality}

Figure~\ref{fig:communication-efficiency} compares message volume with excess
cost. If more communication reliably improved coordination, higher message
counts should predict lower excess cost. The aggregate results do not show that
pattern.

In uniform-cost games, Qwen 3.6 Plus sends the fewest messages (2.96) but has
higher excess cost (0.40) than DeepSeek V4 Pro, which sends more messages
(3.21) and has the lowest excess cost (0.18). Claude Sonnet 4.6 sends the most
messages (3.46) without achieving the lowest excess cost. In varied-cost games,
Llama 4 Maverick is both the least verbose and the lowest-cost model
(3.20 messages; 0.24 excess), but this relationship does not hold generally:
GPT-5.4 Mini sends 3.44 messages and incurs the highest excess cost (1.49),
while DeepSeek V4 Pro sends fewer messages (3.24) and incurs much lower excess
cost (0.38). Raw message count is therefore a weak proxy for coordination
quality; what matters is whether messages convey useful availability and cost
information.

\begin{figure}[ht]
\centering
\includegraphics[width=\linewidth]{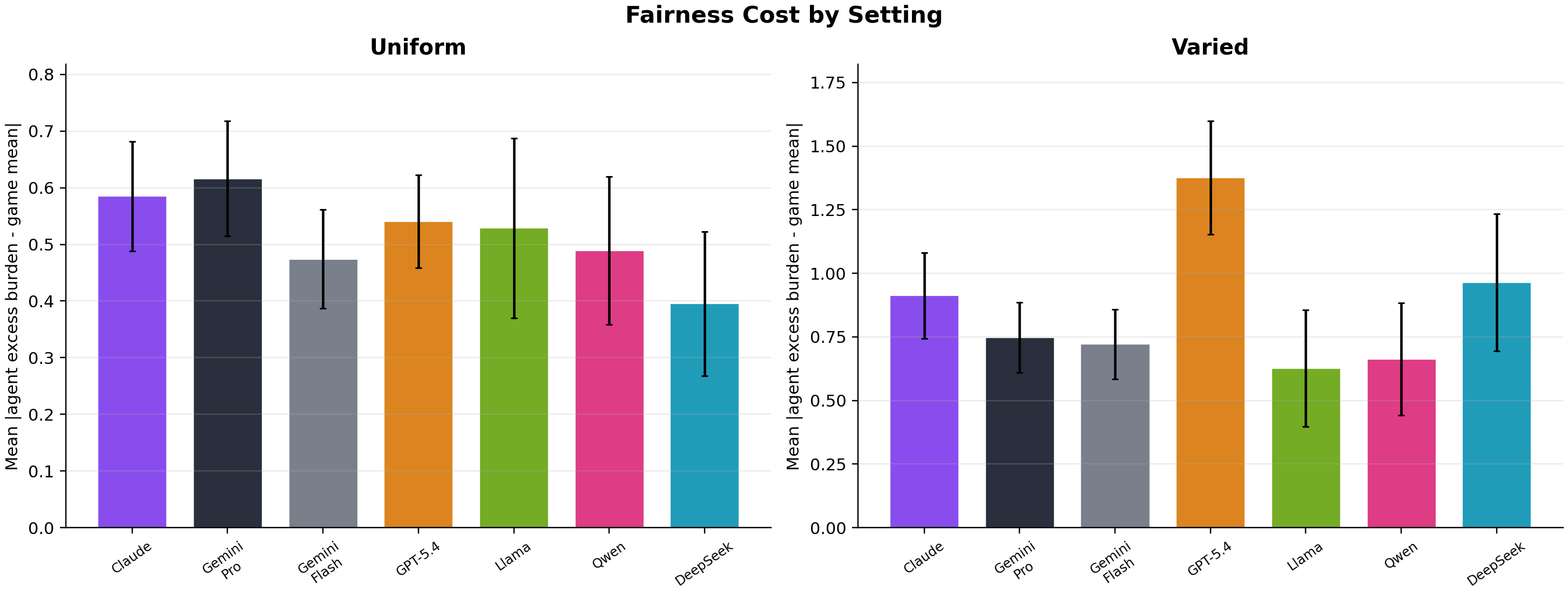}
\caption{Fairness cost by model and cost setting. Lower values mean an agent's
signed excess burden is closer to the within-game average. Error bars show
95\% confidence intervals across task instances.}
\label{fig:fairness-by-setting}
\end{figure}

\paragraph{Fairness in varied-cost games.}
Figure~\ref{fig:fairness-by-setting} shows that fairness costs are tightly
clustered in uniform-cost games, ranging from 0.396 to 0.616. Varied-cost games
create larger burden-allocation differences. Llama 4 Maverick has the lowest
fairness cost (0.627), followed by Qwen 3.6 Plus (0.662) and Gemini 3 Flash
(0.721). GPT-5.4 Mini has the highest fairness cost (1.376).

This suggests a model-specific privacy--fairness tension. GPT-5.4 Mini has low
VPS in varied-cost games (4.60) but the worst fairness cost, whereas Llama 4
Maverick has both the lowest VPS (3.49) and the best fairness. The issue is
therefore not privacy preservation itself, but the kind of information withheld.
A message audit shows that GPT-5.4 Mini proposes its own slots at a similar rate
to other models (0.875 vs.\ 0.866) and reports availability at least as often
(0.853 vs.\ 0.769), but mentions its own costs or constraints far less often
(6.3\% vs.\ 29.2\%; \(p<5\times10^{-6}\)). This suggests that GPT-5.4 Mini's
privacy comes partly from omitting cost-relevant context, leaving teammates with
less evidence for fair burden allocation.

\begin{table}[t]
\centering
\scriptsize
\setlength{\tabcolsep}{6pt}
\caption{Deterministic baseline results on the canonical mixed-agent CalBench task set. Each metric is computed over N = 45 tasks. VPS is computed using the baseline VPS formula and is not directly comparable to the reflection-calibrated VPS reported for language-model agents.}
\label{tab:baseline-results}
\begin{tabular}{lrrrrrr}
\toprule
Method & Coord. & Excess & Msgs. & Fair. & VPS \\
\midrule

\multicolumn{6}{l}{\textbf{Uniform Cost}} \\
IMAP        & 100.0 & 0.26 & 2.00 & 0.530 & 12.40 \\
DSM-private & 45.8  & 0.98 & 4.54 & 0.809 & 0.00  \\
DSM-welfare & 100.0 & 0.27 & 2.69 & 0.530 & 25.05 \\
SD-MAP      & 62.2  & 1.68 & 7.48 & 0.741 & 0.12  \\

\midrule

\multicolumn{6}{l}{\textbf{Varied Cost}} \\
IMAP        & 100.0 & 0.32 & 2.00 & 0.665 & 12.40 \\
DSM-private & 48.9  & 2.08 & 4.43 & 1.865 & 0.00  \\
DSM-welfare & 99.1  & 0.46 & 2.78 & 0.816 & 23.55 \\
SD-MAP      & 63.1  & 3.64 & 7.30 & 1.376 & 0.08  \\

\bottomrule
\end{tabular}
\end{table}

\section{Discussion and Limitations}

The benchmark separates three behaviors that are easy to conflate. First,
booking the meeting is weaker than coordinating well: models can complete all
assigned meetings while incurring extra displacement cost relative to the
scheduled-only oracle. Second, more talk does not reliably reduce that cost; raw
message count is not a quality measure. Third, privacy can conflict with
fairness when agents hide cost-relevant constraints that teammates need for
allocation decisions. 

\textbf{Limitations and Future Work.} The reported results cover one topology
($N=5$, 3 rotating participants); larger groups and longer meeting streams remain
to be tested, though the harness supports them. Reflection-calibrated VPS
measures reported belief movement rather than latent model state. CalBench
assumes truthful reporting, leaving strategic misrepresentation to future work.
The oracle and VPS machinery also
apply to other streaming coordination problems, such as resource allocation or
supply-chain handoffs.

\section{Conclusion}

CalBench tests whether calendar assistants can coordinate from private state
without treating privacy, cost, and fairness as afterthoughts. Its mixed-agent
evaluation shows that feasibility and cost optimization are separable: uniform
games are near-saturated for most models, but varied-cost games expose large
differences in excess burden. Message volume is a poor proxy for coordination
quality, and low VPS can coincide with unfair burden allocation when agents omit
cost-relevant context. CalBench is released with its scenario
generator, reference protocols, VPS pipeline, and trace analysis toolkit.

\bibliography{references}

@article{herbrich2006trueskill,
  title={TrueSkill™: a Bayesian skill rating system},
  author={Herbrich, Ralf and Minka, Tom and Graepel, Thore},
  journal={Advances in neural information processing systems},
  volume={19},
  year={2006}
}

@article{Joshy2024,
    title        = {OpenSkill: A faster asymmetric multi-team, multiplayer rating system},
    author       = {Vivek Joshy},
    year         = 2024,
    journal      = {Journal of Open Source Software},
    publisher    = {The Open Journal},
    volume       = 9,
    number       = 93,
    pages        = 5901,
    doi          = {10.21105/joss.05901},
    url          = {https://doi.org/10.21105/joss.05901}
}

@misc{liu2024agentbench,
      title={AgentBench: Evaluating LLMs as Agents}, 
      author={Xiao Liu and Hao Yu and Hanchen Zhang and Yifan Xu and Xuanyu Lei and Hanyu Lai and Yu Gu and Hangliang Ding and Kaiwen Men and Kejuan Yang and Shudan Zhang and Xiang Deng and Aohan Zeng and Zhengxiao Du and Chenhui Zhang and Sheng Shen and Tianjun Zhang and Yu Su and Huan Sun and Minlie Huang and Yuxiao Dong and Jie Tang},
      year={2025},
      eprint={2308.03688},
      archivePrefix={arXiv},
      primaryClass={cs.AI},
      url={https://arxiv.org/abs/2308.03688}, 
}

@misc{yao2024taubench,
      title={$\tau$-bench: A Benchmark for Tool-Agent-User Interaction in Real-World Domains}, 
      author={Shunyu Yao and Noah Shinn and Pedram Razavi and Karthik Narasimhan},
      year={2024},
      eprint={2406.12045},
      archivePrefix={arXiv},
      primaryClass={cs.AI},
      url={https://arxiv.org/abs/2406.12045}, 
}

@misc{barres2025tau2,
      title={$\tau^2$-Bench: Evaluating Conversational Agents in a Dual-Control Environment}, 
      author={Victor Barres and Honghua Dong and Soham Ray and Xujie Si and Karthik Narasimhan},
      year={2025},
      eprint={2506.07982},
      archivePrefix={arXiv},
      primaryClass={cs.AI},
      url={https://arxiv.org/abs/2506.07982}, 
}

@inproceedings{
zhou2024sotopia,
title={{SOTOPIA}: Interactive Evaluation for Social Intelligence in Language Agents},
author={Xuhui Zhou and Hao Zhu and Leena Mathur and Ruohong Zhang and Haofei Yu and Zhengyang Qi and Louis-Philippe Morency and Yonatan Bisk and Daniel Fried and Graham Neubig and Maarten Sap},
booktitle={The Twelfth International Conference on Learning Representations},
year={2024},
url={https://openreview.net/forum?id=mM7VurbA4r}
}

@misc{zhu2025multiagentbench,
      title={MultiAgentBench: Evaluating the Collaboration and Competition of LLM agents}, 
      author={Kunlun Zhu and Hongyi Du and Zhaochen Hong and Xiaocheng Yang and Shuyi Guo and Zhe Wang and Zhenhailong Wang and Cheng Qian and Xiangru Tang and Heng Ji and Jiaxuan You},
      year={2025},
      eprint={2503.01935},
      archivePrefix={arXiv},
      primaryClass={cs.MA},
      url={https://arxiv.org/abs/2503.01935}, 
}

@misc{cooperbench2026,
      title={CooperBench: Why Coding Agents Cannot be Your Teammates Yet}, 
      author={Arpandeep Khatua and Hao Zhu and Peter Tran and Arya Prabhudesai and Frederic Sadrieh and Johann K. Lieberwirth and Xinkai Yu and Yicheng Fu and Michael J. Ryan and Jiaxin Pei and Diyi Yang},
      year={2026},
      eprint={2601.13295},
      archivePrefix={arXiv},
      primaryClass={cs.LG},
      url={https://arxiv.org/abs/2601.13295}, 
}

@misc{davidson2025collaborationgap,
      title={The Collaboration Gap}, 
      author={Tim R. Davidson and Adam Fourney and Saleema Amershi and Robert West and Eric Horvitz and Ece Kamar},
      year={2025},
      eprint={2511.02687},
      archivePrefix={arXiv},
      primaryClass={cs.AI},
      url={https://arxiv.org/abs/2511.02687}, 
}

@misc{vink2025reproducibility,
      title={Reproducibility Study of Cooperation, Competition, and Maliciousness: LLM-Stakeholders Interactive Negotiation}, 
      author={Jose L. Garcia and Karolina Hajkova and Maria Marchenko and Carlos Miguel Pati\~{n}o},
      year={2025},
      eprint={2502.16242},
      archivePrefix={arXiv},
      primaryClass={cs.AI},
      url={https://arxiv.org/abs/2502.16242}, 
}

@misc{nath2026craft,
      title={CRAFT: Grounded Multi-Agent Coordination Under Partial Information}, 
      author={Abhijnan Nath and Hannah VanderHoeven and Nikhil Krishnaswamy},
      year={2026},
      eprint={2603.25268},
      archivePrefix={arXiv},
      primaryClass={cs.CL},
      url={https://arxiv.org/abs/2603.25268}, 
}

@inproceedings{lewis2017deal,
    title = "Deal or No Deal? End-to-End Learning of Negotiation Dialogues",
    author = "Lewis, Mike  and
      Yarats, Denis  and
      Dauphin, Yann  and
      Parikh, Devi  and
      Batra, Dhruv",
    editor = "Palmer, Martha  and
      Hwa, Rebecca  and
      Riedel, Sebastian",
    booktitle = "Proceedings of the 2017 Conference on Empirical Methods in Natural Language Processing",
    month = sep,
    year = "2017",
    address = "Copenhagen, Denmark",
    publisher = "Association for Computational Linguistics",
    url = "https://aclanthology.org/D17-1259/",
    doi = "10.18653/v1/D17-1259",
    pages = "2443--2453"
}

@article{fair2022diplomacy,
  title = {Human-level play in the game of {Diplomacy} by combining language models with strategic reasoning},
  author = {{Meta Fundamental AI Research Diplomacy Team (FAIR)} and Bakhtin, Anton and Brown, Noam and Dinan, Emily and Farina, Gabriele and Flaherty, Colin and Fried, Daniel and Goff, Andrew and Gray, Jonathan and Hu, Hengyuan and Jacob, Athul Paul and Komeili, Mojtaba and Konath, Karthik and Kwon, Minae and Lerer, Adam and Lewis, Mike and Miller, Alexander H. and Mitts, Sasha and Renduchintala, Adithya and Roller, Stephen and Rowe, Dirk and Shi, Weiyan and Spisak, Joe and Wei, Alexander and Wu, David and Zhang, Hugh and Zijlstra, Markus},
  journal = {Science},
  volume = {378},
  number = {6624},
  pages = {1067--1074},
  year = {2022},
  doi = {10.1126/science.ade9097}
}

@inproceedings{bianchi2024negotiationarena,
author = {Bianchi, Federico and Chia, Patrick John and Yuksekgonul, Mert and Tagliabue, Jacopo and Jurafsky, Dan and Zou, James},
title = {How well can LLMs negotiate? NEGOTIATIONARENA platform and analysis},
year = {2024},
publisher = {JMLR.org},
booktitle = {Proceedings of the 41st International Conference on Machine Learning},
articleno = {158},
numpages = {17},
location = {Vienna, Austria},
series = {ICML'24}
}

@inproceedings{abdelnabi2024llmdeliberation,
author = {Abdelnabi, Sahar and Gomaa, Amr and Sivaprasad, Sarath and Sch\"{o}nherr, Lea and Fritz, Mario},
title = {Cooperation, competition, and maliciousness: LLM-stakeholders interactive negotiation},
year = {2024},
isbn = {9798331314385},
publisher = {Curran Associates Inc.},
address = {Red Hook, NY, USA},
booktitle = {Proceedings of the 38th International Conference on Neural Information Processing Systems},
articleno = {2658},
numpages = {52},
location = {Vancouver, BC, Canada},
series = {NIPS '24}
}

@inproceedings{maheswaran2004dimes,
  title = {Taking {DCOP} to the Real World: Efficient Complete Solutions for Distributed Multi-Event Scheduling},
  author = {Maheswaran, Rajiv T. and Tambe, Milind and Bowring, Emma and Pearce, Jonathan P. and Varakantham, Pradeep},
  booktitle = {Proceedings of the Third International Joint Conference on Autonomous Agents and Multiagent Systems},
  series = {AAMAS '04},
  pages = {310--317},
  year = {2004},
  publisher = {IEEE Computer Society},
  address = {Washington, DC, USA},
  doi = {10.1109/AAMAS.2004.257},
  isbn = {9781581138641}
}

@article{maheswaran2006vps,
  title = {Privacy Loss in Distributed Constraint Reasoning: A Quantitative Framework for Analysis and its Applications},
  author = {Maheswaran, Rajiv T. and Pearce, Jonathan P. and Bowring, Emma and Varakantham, Pradeep and Tambe, Milind},
  journal = {Autonomous Agents and Multi-Agent Systems},
  volume = {13},
  pages = {27--60},
  year = {2006},
  doi = {10.1007/s10458-006-5951-y},
  publisher = {Springer}
}

@inproceedings{greenstadt2006privacy,
  title = {Analysis of Privacy Loss in Distributed Constraint Optimization},
  author = {Greenstadt, Rachel and Pearce, Jonathan P. and Tambe, Milind},
  booktitle = {Proceedings of the Twenty-First National Conference on Artificial Intelligence},
  series = {AAAI'06},
  pages = {647--653},
  year = {2006},
  publisher = {AAAI Press}
}

@inproceedings{brito2008privacy,
  title = {Privacy in Distributed Meeting Scheduling},
  author = {Brito, Ismel and Meseguer, Pedro},
  booktitle = {Frontiers in Artificial Intelligence and Applications},
  volume = {184},
  pages = {118--127},
  year = {2008},
  doi = {10.3233/978-1-58603-925-7-118}
}

@article{enembreck2012mulbs,
  title = {Distributed Constraint Optimization with {MULBS}: A Case Study on Collaborative Meeting Scheduling},
  author = {Enembreck, Fabr{\'i}cio and Barth{\`e}s, Jean-Paul Andr{\'e}},
  journal = {Journal of Network and Computer Applications},
  volume = {35},
  number = {1},
  pages = {164--175},
  year = {2012},
  doi = {10.1016/j.jnca.2011.02.016}
}

@article{nissenbaum2004ci,
  title = {Privacy as Contextual Integrity},
  author = {Nissenbaum, Helen},
  journal = {Washington Law Review},
  volume = {79},
  number = {1},
  pages = {119--158},
  year = {2004}
}

@misc{mireshghallah2024confaide,
      title={Can LLMs Keep a Secret? Testing Privacy Implications of Language Models via Contextual Integrity Theory}, 
      author={Niloofar Mireshghallah and Hyunwoo Kim and Xuhui Zhou and Yulia Tsvetkov and Maarten Sap and Reza Shokri and Yejin Choi},
      year={2024},
      eprint={2310.17884},
      archivePrefix={arXiv},
      primaryClass={cs.AI},
      url={https://arxiv.org/abs/2310.17884}, 
}

@misc{shao2024privacylens,
      title={PrivacyLens: Evaluating Privacy Norm Awareness of Language Models in Action}, 
      author={Yijia Shao and Tianshi Li and Weiyan Shi and Yanchen Liu and Diyi Yang},
      year={2025},
      eprint={2409.00138},
      archivePrefix={arXiv},
      primaryClass={cs.CL},
      url={https://arxiv.org/abs/2409.00138}, 
}

@inproceedings{bagdasaryan2024airgap,
author = {Bagdasarian, Eugene and Yi, Ren and Ghalebikesabi, Sahra and Kairouz, Peter and Gruteser, Marco and Oh, Sewoong and Balle, Borja and Ramage, Daniel},
title = {AirGapAgent: Protecting Privacy-Conscious Conversational Agents},
year = {2024},
isbn = {9798400706363},
publisher = {Association for Computing Machinery},
address = {New York, NY, USA},
url = {https://doi.org/10.1145/3658644.3690350},
doi = {10.1145/3658644.3690350},
booktitle = {Proceedings of the 2024 on ACM SIGSAC Conference on Computer and Communications Security},
pages = {3868--3882},
numpages = {15},
location = {Salt Lake City, UT, USA},
series = {CCS '24}
}

@inproceedings{debenedetti2024agentdojo,
  title = {{AgentDojo}: A Dynamic Environment to Evaluate Prompt Injection Attacks and Defenses for {LLM} Agents},
  author = {Debenedetti, Edoardo and Zhang, Jie and Balunovic, Mislav and Beurer-Kellner, Luca and Fischer, Marc and Tram{\`e}r, Florian},
  booktitle = {Advances in Neural Information Processing Systems},
  volume = {37},
  pages = {82895--82920},
  year = {2024},
  publisher = {Curran Associates, Inc.},
  doi = {10.52202/079017-2636},
  url = {https://proceedings.neurips.cc/paper_files/paper/2024/file/97091a5177d8dc64b1da8bf3e1f6fb54-Paper-Datasets_and_Benchmarks_Track.pdf}
}

@misc{elyagoubi2026agentleak,
      title={AgentLeak: A Full-Stack Benchmark for Privacy Leakage in Multi-Agent LLM Systems}, 
      author={Faouzi El Yagoubi and Godwin Badu-Marfo and Ranwa Al Mallah},
      year={2026},
      eprint={2602.11510},
      archivePrefix={arXiv},
      primaryClass={cs.AI},
      url={https://arxiv.org/abs/2602.11510}, 
}

@misc{magpie2025,
      title={MAGPIE: A dataset for Multi-AGent contextual PrIvacy Evaluation}, 
      author={Gurusha Juneja and Alon Albalak and Wenyue Hua and William Yang Wang},
      year={2025},
      eprint={2506.20737},
      archivePrefix={arXiv},
      primaryClass={cs.AI},
      url={https://arxiv.org/abs/2506.20737}, 
}

@inproceedings{
renting2024meeting,
title={Multi-Agent Meeting Scheduling: A Negotiation Perspective},
author={Bram M. Renting and Holger Hoos and Catholijn M Jonker},
booktitle={The Sixteenth Workshop on Adaptive and Learning Agents},
year={2024},
url={https://openreview.net/forum?id=jNmtve2Av2}
}

@article{farhadi2021faithful,
author = {Farhadi, Farzaneh and Jennings, Nicholas R.},
title = {A Faithful Mechanism for Incremental Multi-Agent Agreement Problems with Self-Interested and Privacy-Preserving Agents},
year = {2021},
issue_date = {Jul 2021},
publisher = {Springer-Verlag},
address = {Berlin, Heidelberg},
volume = {2},
number = {4},
url = {https://doi.org/10.1007/s42979-021-00650-4},
doi = {10.1007/s42979-021-00650-4},
journal = {SN Comput. Sci.},
month = may,
numpages = {33}
}

@inproceedings{
mahmud2025collab,
title={Co{LLAB}: A Framework for Designing Scalable Benchmarks for Agentic {LLM}s},
author={Saaduddin Mahmud and Eugene Bagdasarian and Shlomo Zilberstein},
booktitle={Workshop on Scaling Environments for Agents},
year={2025},
url={https://openreview.net/forum?id=372FjQy1cF}
}

@inproceedings{modi2004multiagent,
  title     = {Multiagent Meeting Scheduling with Rescheduling},
  author    = {Modi, Pragnesh Jay and Veloso, Manuela},
  booktitle = {Proceedings of the Fifth Workshop on Distributed Constraint Reasoning (DCR)},
  year      = {2004}
}

@misc{zheng2024naturalplan,
      title={NATURAL PLAN: Benchmarking LLMs on Natural Language Planning}, 
      author={Huaixiu Steven Zheng and Swaroop Mishra and Hugh Zhang and Xinyun Chen and Minmin Chen and Azade Nova and Le Hou and Heng-Tze Cheng and Quoc V. Le and Ed H. Chi and Denny Zhou},
      year={2024},
      eprint={2406.04520},
      archivePrefix={arXiv},
      primaryClass={cs.CL},
      url={https://arxiv.org/abs/2406.04520}, 
}

@article{crawford1982strategic,
  title = {Strategic Information Transmission},
  author = {Crawford, Vincent P. and Sobel, Joel},
  journal = {Econometrica},
  volume = {50},
  number = {6},
  pages = {1431--1451},
  year = {1982},
  doi = {10.2307/1913390},
  publisher = {The Econometric Society}
}

@misc{cpsatlp,
  title = {CP-SAT},
  author = {Laurent Perron and Fr{\'e}d{\'e}ric Didier},
  year = {2025},
  url = {https://developers.google.com/optimization/cp/cp_solver/},
  note = {Google OR-Tools, version 9.12}
}

@misc{microsoft2026socialreasoning,
  title = {SocialReasoning-Bench: Measuring whether AI agents act in users' best interests},
  author = {Payne, Tyler and Epperson, Will and Yousefi, Safoora and Huang, Zachary and Bansal, Gagan and Hua, Wenyue and Murad, Maya and Celikyilmaz, Asli and Amershi, Saleema},
  year = {2026},
  month = may,
  howpublished = {Microsoft Research Blog},
  url = {https://www.microsoft.com/en-us/research/blog/socialreasoning-bench-measuring-whether-ai-agents-act-in-users-best-interests/}
}

\appendix


\section{Prompt Construction and Game Scaffolding}
\label{app:scaffolding}
\label{app:prompts}

This appendix provides a self-contained description of how the experiment
harness constructs agent prompts, sequences messages across game phases, and
validates agent responses.  Together with the system-prompt text in
\S\ref{app:prompts} and the trace format in \S\ref{app:scaffolding-trace},
this section contains sufficient detail to replicate the communication
protocol from scratch.

\subsection{Scenario and Calendar Generation}
\label{app:scaffolding-scenario}

Every game is seeded from a \emph{scenario}, a deterministic data structure
produced by \texttt{generate\_scenario(seed, \ldots)}.  Fixing the seed
reproduces the exact calendar layouts, cost functions, participant lists, and
witness solution used in the original run.

\paragraph{Parameters.}
Table~\ref{tab:scenario-params} lists the key generation parameters.

\begin{table*}[t]
\centering\small
\caption{Scenario generation parameters.}
\label{tab:scenario-params}
\begin{tabular}{lll}
\toprule
\textbf{Parameter} & \textbf{Type} & \textbf{Description} \\
\midrule
\texttt{seed}              & int   & RNG seed; fixes all randomness \\
\texttt{num\_agents}       & int   & Number of agents $N$ \\
\texttt{num\_slots}        & int   & Calendar length $S$ (default 16) \\
\texttt{density}           & float or list & Shared or per-agent fraction of slots pre-filled with errands, $\in[0,1]$ \\
\texttt{num\_meetings}     & int   & Number of rounds $C$ \\
\texttt{pref\_level}       & int   & Max errand cost draw from $\text{Uniform}[1, \texttt{pref\_level}]$ \\
\texttt{errand\_cost\_level}& int  & Override for errand cost ceiling (defaults to \texttt{pref\_level}) \\
\texttt{meeting\_cost\_level}& int & Max prior-meeting cost \\
\texttt{participant\_lists}& list  & Optional fixed participant sets per round \\
\bottomrule
\end{tabular}
\end{table*}

\paragraph{Calendar construction.}
The generator uses a backward-from-witness design to guarantee feasibility:

\begin{enumerate}[leftmargin=1.5em]
  \item A \emph{witness slot} $w_c$ is drawn uniformly at random for each
        meeting $c$ from slots not already used by any of its participants.
        This hidden solution achieves a known cost.
  \item For each agent $i$, errand items are placed in $\lfloor S \cdot
        \rho_i \rfloor$ slots drawn randomly, where $\rho_i$ is either the
        shared \texttt{density} value or the agent-specific entry in a density
        vector.  Witness slots are deliberately seeded with errands (when
        \texttt{force\_witness\_errand} is true) to create displacement
        pressure.
  \item A small set of \emph{absorbing slots} --- one per errand occupying a
        witness slot --- is kept free so displaced errands always have a valid
        landing pad.  This preserves the invariant that the witness solution
        is feasible under any sequence of moves needed to clear the witness
        slots.
  \item Each errand is assigned a displacement cost drawn from
        $\text{Uniform}[1, \texttt{errand\_cost\_level}]$.
  \item The optimal cost, greedy cost, and CP-SAT feasible-assignment fraction
        are computed and stored in the scenario dict.  The feasible-assignment
        fraction counts distinct injective meeting-slot assignments that satisfy
        the CP-SAT constraints and divides by $S!/(S-C)!$.
\end{enumerate}

\paragraph{Calendar data model.}
Each agent's calendar is a fixed-length list of \emph{slots}.  A slot is
either \texttt{null} (free), an errand object
\texttt{\{errand\_id, cost\}}, or a meeting object
\texttt{\{meeting\_id, cost\}}.  Blocked errands carry an additional
\texttt{blocked: true} flag and cannot be moved.  The rendered string shown
to LLM agents is produced by \texttt{Calendar.render()}, which emits one line
per slot in the format:

\begin{promptbox}[Calendar Render]
Slot  0: [FREE]\\
Slot  1: Errand \#3 (cost=2)\\
Slot  2: Blocked Errand \#7 (cost=5)\\
Slot  3: Meeting M2 (cost=1) participants=[0, 2]
\end{promptbox}

\paragraph{Private label hydration.}
Before the rendered calendar or meeting description is delivered to an agent,
the harness injects \emph{private semantic labels} --- naturalistic event
descriptions such as ``Intake session at the Willow Tree Eating Disorder
Clinic'' --- drawn from a pre-generated label bank.  Labels are assigned to
slots once per scenario and remain stable across all rounds.  The hydration
layer enforces information asymmetry:

\begin{itemize}[leftmargin=1.5em]
  \item An agent sees its own errand label on every turn.
  \item An agent sees a meeting's private label only if it is a participant
        in that meeting.
  \item Non-participant agents see the slot type (\texttt{errand},
        \texttt{meeting}) and cost but not the semantic label.
\end{itemize}

\noindent This asymmetry is enforced at the scaffolding level and does not
depend on the LLM's own compliance with privacy instructions.

\paragraph{Label bank construction.}
Labels are generated offline using an LLM (Gemini 3 Flash) prompted with one
of three tier-specific instructions: \textsc{sensitive} (medical, legal,
financial, identity), \textsc{neutral} (ambiguously private everyday
logistics), and \textsc{public} (routine low-stakes errands).  The resulting
label bank is stored as a JSON file and loaded deterministically at scenario
generation time.

\subsection{System Prompt Construction}
\label{app:scaffolding-system}
\label{app:prompt-standard}

Each agent receives a single \textbf{system prompt} at registration time,
delivered as the \texttt{system} role message before any user turns.  The
prompt is assembled once per agent from a fixed template parameterised by the
following game-configuration values:

\begin{itemize}[leftmargin=1.5em]
  \item \texttt{agent\_id} --- the agent's own integer identifier.
  \item \texttt{all\_agent\_ids} --- the complete list of agent identifiers in
        the game (e.g.\ \texttt{0, 1, 2, 3, 4} for a 5-agent game).
  \item \texttt{num\_agents} --- total number of agents.
  \item \texttt{num\_slots} --- number of calendar slots.
  \item \texttt{decision\_retries} --- maximum retries allowed if a DECISION
        batch fails validation.
\end{itemize}

\noindent The template is divided into nine fixed sections (shown verbatim in
\S\ref{app:prompt-standard}):

\begin{enumerate}[leftmargin=1.5em]
  \item \textsc{Rules} --- hard constraints: no double-booking, same-slot
        consistency for all participants, same-slot constraint for
        rescheduled meetings.
  \item \textsc{Important Constraints} --- information constraints: own-only
        visibility, prohibition on sharing errand/meeting details with
        non-participants, prohibition on sharing cost functions.
  \item \textsc{Negotiation Strategy} --- six negotiation guidelines
        instructing the agent to prefer free slots, push back on costly
        proposals, and use only qualitative language about difficulty.
  \item \textsc{Calendar Slot Types} --- definitions of the four slot types
        (\texttt{free}, \texttt{blocked}, \texttt{errand}, \texttt{meeting}).
  \item \textsc{Tools} --- the JSON tool-call schema for \texttt{dm},
        participant groupchat, all-agent groupchat, \texttt{schedule}, and
        \texttt{reschedule}, with phase-specific validity rules.
  \item \textsc{Phases} --- descriptions of the four phases each agent
        participates in (CHEAP\_TALK, VOLUNTARY, DECISION, RESOLUTION).
  \item \textsc{Response Format} --- the required \texttt{\{"thinking": ...,
        "actions": [...]\}} envelope with a worked example.
  \item \textsc{Identity} --- the agent's own ID and the full agent list.
  \item \textsc{Environment Parameters} --- slot count, retry budget,
        multi-round structure.
\end{enumerate}

\paragraph{Adversarial variant injection.}
\label{app:prompt-nosy}
\label{app:prompt-redteam}
Agents configured with the \textsc{nosy-highpressure} or \textsc{redteam}
variants receive one additional section appended verbatim after the base
prompt (see \S\ref{app:prompt-nosy} and \S\ref{app:prompt-redteam}).  No
other section changes; the injection is self-contained.

\paragraph{DSPy variant injection.}
Agents configured with a DSPy prompt variant (\texttt{type: dspy} in the
experiment YAML) receive a separately versioned Markdown block loaded from a
\texttt{prompt\_variants/} directory and appended after the base prompt.  The
variant filename and directory are stored in the experiment config and trace
metadata, making the augmentation fully reproducible.  The base prompt is
identical to the standard agent prompt; the DSPy block is the only difference.

\subsection{Per-Round User Message Sequence}
\label{app:scaffolding-messages}

After registration, the harness drives each round by appending \textbf{user-role
messages} to each agent's conversation.  A single game keeps one
per-agent conversation thread alive for the full multi-round game; messages
accumulate in the thread rather than being reset between rounds.
Table~\ref{tab:message-sequence} lists the message types in delivery order.
The same round loop is used for all communication conditions.  The private-DM
condition exposes only addressed messages, the meeting-participant groupchat
condition exposes a shared channel to the current meeting participants, and the
all-agent groupchat condition exposes a shared channel to every agent.  Meeting
participants are active by default; non-participants are activated only by a
direct DM or by an all-agent groupchat message.

\begin{table*}[t]
\centering\small
\caption{User-message sequence per round.  Each row is one message appended
to an agent's conversation thread.}
\label{tab:message-sequence}
\begin{tabular}{p{3.5cm}p{4.5cm}p{3.5cm}}
\toprule
\textbf{Phase / condition} & \textbf{Message builder} & \textbf{Delivered to} \\
\midrule
CHEAP\_TALK, turn 0
  & \texttt{build\_round\_start\_message}
  & all participants \\
CHEAP\_TALK, turn $t \geq 1$, agent received messages
  & \texttt{build\_turn\_message}
  & agents with non-empty inbox \\
CHEAP\_TALK, turn $t \geq 1$, no new messages
  & \texttt{build\_turn\_message} (empty inbox path)
  & agents with empty inbox but still in speaker order \\
VOLUNTARY (non-participant contacted)
  & \texttt{build\_voluntary\_reschedule\_message}
  & non-participants activated during CHEAP\_TALK \\
DECISION
  & \texttt{build\_decision\_message}
  & all participants \\
DECISION retry (batch rejected)
  & \texttt{build\_retry\_message}
  & the participant whose batch failed \\
\bottomrule
\end{tabular}
\end{table*}

\paragraph{Round-start message.}
Produced by \texttt{build\_round\_start\_message(meeting, calendar\_render,
round\_num, incurred\_penalty, turn\_index, max\_turns\_per\_round)}.
Delivered to every participant at the start of CHEAP\_TALK (turn 0).
Contains, in order:

\begin{enumerate}[leftmargin=1.5em]
  \item A \texttt{=== ROUND N START ===} header.
  \item The meeting to schedule: ID, participant list, duration in slots, and
        the private meeting label (visible only to participants).
  \item The agent's own calendar as a rendered slot list.
  \item The agent's cumulative penalty incurred in previous rounds.
  \item The active phase (\texttt{CHEAP\_TALK}) and, when
        \texttt{max\_turns\_per\_round} is set, a turn-budget line of the
        form ``CHEAP\_TALK turn budget: turn 1 of N.  $k$ turn(s) remain
        after this one.''
  \item A reminder that DECISION follows CHEAP\_TALK and that agents should
        negotiate for a low-displacement slot.
\end{enumerate}

\paragraph{Turn message.}
Produced by \texttt{build\_turn\_message(messages, turn\_index,
max\_turns\_per\_round)}.  Delivered for every subsequent CHEAP\_TALK turn to
agents in the current speaker order.  When the agent has new messages, the
body lists each incoming private or groupchat message with channel metadata,
for example:
\begin{promptbox}[Incoming Message]
[1] From Agent 2 (meeting 3): <content>
\end{promptbox}
When the inbox is empty, the body reads ``No new messages in your inbox.''
On the final allowed turn (\texttt{turn\_index + 1 == max\_turns\_per\_round}),
an additional sentence instructs the agent not to ask open-ended questions and
to return \texttt{[]} if coordination is complete.

\paragraph{Voluntary-reschedule message.}
Produced by \texttt{build\_voluntary\_reschedule\_message(meeting,
calendar\_render)}.  Delivered after CHEAP\_TALK closes to non-participant
agents who received at least one DM during the round.  The message explains
that the agent may execute \texttt{reschedule} calls to honour commitments
made during negotiation but may not use \texttt{schedule}.

\paragraph{Decision message.}
Produced by \texttt{build\_decision\_message(meeting, calendar\_render)}.
Delivered to each participant to collect the final scheduling batch.  The
message repeats the meeting metadata and a \emph{frozen} calendar snapshot
taken at the moment DECISION begins (post-VOLUNTARY).  It instructs the agent
to submit at least one \texttt{schedule} call and any \texttt{reschedule}
calls needed to free the target slot, reminds the agent that the batch is
resolved atomically, and instructs the agent to use the slot agreed during
CHEAP\_TALK.

\paragraph{Retry message.}
Produced by \texttt{build\_retry\_message(attempt, max\_attempts, conflict)}.
Delivered when a DECISION batch fails validation.  The body states the
attempt counter, the exact conflict description (e.g.\ ``Expected exactly 1
schedule action, got 0''), and instructs the agent to resubmit the complete
batch from scratch without referencing the previous attempt.

\subsection{Game-Round Phase Orchestration}
\label{app:scaffolding-phases}

The harness runs each round synchronously within a single async loop.
Algorithm~\ref{alg:round} gives the complete pseudocode.

\begin{algorithm*}[t]
\small
\caption{One game round in the calendar scheduling harness.}
\label{alg:round}
\begin{algorithmic}[1]
\Require meeting $m$, speaker order $P$ (participants), all agents $A$,
         \texttt{max\_turns}
\State Emit \textsc{round\_start} event
\State \textbf{// CHEAP\_TALK PHASE}
\State $t \gets 0$; \quad $\textit{has\_activity} \gets \textsc{true}$
\While{$\textit{has\_activity}$ \textbf{and} $t < \texttt{max\_turns}$}
  \State $\textit{has\_activity} \gets \textsc{false}$
  \ForAll{$a \in P$ in speaker order}
    \State msg $\gets$ \textsc{RoundStart}($m$, cal$_a$, $t$) \textbf{if} $t{=}0$ \textbf{else} \textsc{Turn}(inbox$_a$, $t$)
    \State Deliver msg; emit \textsc{turn\_start}
    \State result $\gets a.\texttt{turn}(t)$; emit \textsc{turn\_end}
    \ForAll{message $d \in$ result.tool\_calls}
      \State Route $d$ to visible recipients; emit message event
      \State $\textit{has\_activity} \gets \textsc{true}$
      \If{$d$ activates non-participants} queue those agents \EndIf
    \EndFor
  \EndFor
  \ForAll{non-participant $a$ queued this turn}
    \State Deliver \textsc{Turn}(inbox$_a$, $t$); emit \textsc{turn\_start}
    \State result $\gets a.\texttt{turn}(t)$; emit \textsc{turn\_end}
    \State Process any DMs in result (same as above)
  \EndFor
  \State $t \gets t + 1$
\EndWhile
\State \textbf{// VOLUNTARY PHASE}
\ForAll{non-participant $a$ activated during CHEAP\_TALK}
  \State Deliver \textsc{Voluntary}($m$, cal$_a$); emit \textsc{decide\_start}
  \State result $\gets a.\texttt{voluntary\_decide}(m)$; emit \textsc{decide\_end}
  \State actions $\gets$ [calls with type \texttt{reschedule}]
  \For{attempt $= 0$ to \texttt{decision\_retries}}
    \State (ok, err) $\gets$ \textsc{ValidateBatch}(cal$_a$, actions, \textit{require\_schedule}$=\textsc{false}$)
    \If{ok} apply actions to cal$_a$; emit \textsc{batch\_applied}; \textbf{break}
    \Else\ emit \textsc{batch\_rejected}; result $\gets a.\texttt{retry\_decide}(\cdot)$
    \EndIf
  \EndFor
\EndFor
\State \textbf{// DECISION PHASE}
\ForAll{$a \in P$}
  \State Deliver \textsc{Decision}($m$, cal$_a$); emit \textsc{decide\_start}
  \State result $\gets a.\texttt{decide}(m)$; emit \textsc{decide\_end}
  \State actions $\gets$ result.tool\_calls
  \For{attempt $= 0$ to \texttt{decision\_retries}}
    \State (ok, err) $\gets$ \textsc{ValidateBatch}(cal$_a$, actions, \textit{require\_schedule}$=\textsc{true}$)
    \If{ok} stage (cal$_a$, actions, cost); \textbf{break}
    \Else\ emit \textsc{batch\_rejected}; Deliver \textsc{Retry}(attempt, err); result $\gets a.\texttt{decide}(m)$
    \EndIf
  \EndFor
\EndFor
\State Commit all staged decisions atomically
\State \textbf{// RESOLUTION PHASE}
\State $\textit{slots} \gets \{$scheduled slot chosen by each $a \in P\}$
\If{$|\textit{slots}| = 1$} round succeeds; update game state
\Else\ record consistency violation; do not update
\EndIf
\end{algorithmic}
\end{algorithm*}

\paragraph{Message routing.}
A private DM is a tool call \texttt{\{"type": "dm", "to": <int>, "content":
"<string>"\}} emitted during CHEAP\_TALK.  The meeting-participant groupchat
and all-agent groupchat tools use the same content payload but differ in
visibility: participant groupchat messages are appended to the inboxes of the
current meeting participants, while all-agent groupchat messages are appended
to every agent's inbox.  Recipients drain their inboxes at the start of their
next turn; the drained snapshot is what populates the \textsc{Turn} message.
Non-participant agents who receive a private DM or all-agent groupchat message
are queued to speak in the same CHEAP\_TALK turn and may themselves send
messages, which triggers further routing.  Participant-only groupchat does not
activate non-participants.  Each turn index $t$ corresponds to one sweep through
all active speakers.

\paragraph{Turn budget.}
When \texttt{max\_turns\_per\_round} is set, the loop terminates after that
many sweeps even if DMs are still being sent.  The final turn message
includes an explicit instruction to close coordination, preventing unbounded
negotiation.

\paragraph{Participant vs.\ non-participant paths.}
At each CHEAP\_TALK turn, participants always receive a message (round-start
on turn 0, turn message thereafter) regardless of whether they have inbox
messages.  Non-participants only receive a message if they were queued because
a direct DM or all-agent groupchat message reached them; otherwise they are
silent.

\subsection{Batch Validation Rules}
\label{app:scaffolding-validation}

Every agent response must be a JSON object with exactly two keys:

\begin{itemize}[leftmargin=1.5em]
  \item \texttt{"thinking"} --- a free-text chain-of-thought string.  Logged
        in the trace but never forwarded to other agents.
  \item \texttt{"actions"} --- a list of tool-call objects, or \texttt{[]} to
        pass without action.
\end{itemize}

\noindent The harness extracts \texttt{"actions"} and applies the following
validation rules before committing any changes to the calendar:

\begin{enumerate}[leftmargin=1.5em]
  \item \textbf{Slot bounds.}  All \texttt{slot}, \texttt{from\_slot}, and
        \texttt{to\_slot} values must be integers in $[0, S)$.
  \item \textbf{Item identity.}  Each \texttt{reschedule} action's
        \texttt{item\_id} must match the \texttt{errand\_id} or
        \texttt{meeting\_id} of the item actually present at
        \texttt{from\_slot}.
  \item \textbf{No blocked moves.}  Items with \texttt{blocked: true} may
        not be moved.
  \item \textbf{No destination conflicts.}  No two actions in the batch may
        target the same \texttt{to\_slot}.
  \item \textbf{Freeness after batch.}  Each \texttt{to\_slot} must be free
        on the calendar \emph{or} be freed by another \texttt{reschedule}
        in the same batch.
  \item \textbf{Exactly one schedule.}  In the DECISION phase, the batch
        must contain exactly one \texttt{schedule} action.  (The VOLUNTARY
        phase uses \textit{require\_schedule}$=\textsc{false}$.)
  \item \textbf{Schedule slot free.}  The \texttt{schedule} slot must be
        free after all \texttt{reschedule} operations are applied.
\end{enumerate}

\noindent Validation is transactional: nothing is applied until the entire
batch passes.  On failure the harness emits a \textsc{batch\_rejected} event
with the exact conflict string and delivers a \textsc{Retry} message to the
agent.  Up to \texttt{decision\_retries} retries are attempted; if all fail,
the last attempted batch is discarded and the round proceeds without that
agent's scheduling action, triggering a consistency violation.

\subsection{Trace Format}
\label{app:scaffolding-trace}

Each game run produces a single JSON file containing:

\begin{itemize}[leftmargin=1.5em]
  \item \texttt{game\_id} --- UUID generated at run time.
  \item \texttt{config} --- the full experiment configuration dict, including
        all agent specs, scenario parameters, and experiment metadata.
  \item \texttt{events} --- a chronologically ordered list of typed event
        objects.  Key event types are listed in Table~\ref{tab:event-types}.
  \item \texttt{final\_state} --- aggregate metrics: total displacement cost,
        rounds succeeded/failed, DM counts, consistency violations.
  \item \texttt{metrics} --- derived metrics computed at run end (e.g.\
        normalised cost vs.\ optimal, cost vs.\ greedy).
  \item \texttt{started\_at}, \texttt{ended\_at} --- ISO-8601 timestamps.
\end{itemize}

\begin{table*}[t]
\centering\small
\caption{Key event types in the game trace.}
\label{tab:event-types}
\begin{tabular}{lp{8cm}}
\toprule
\textbf{Event type} & \textbf{Contents} \\
\midrule
\texttt{game\_start}      & Scenario seed, num agents/slots, optimal/greedy cost, nosy agent IDs \\
\texttt{agent\_registered}& Agent ID, system prompt text, initial calendar render, is-nosy flag \\
\texttt{round\_start}     & Round number, meeting dict, speaker order \\
\texttt{turn\_start}      & Round, turn, phase, agent ID, inbox snapshot, prompt sent \\
\texttt{turn\_end}        & Tool calls, thinking text, token usage, latency \\
\texttt{dm\_sent}         & From/to agent IDs, meeting ID, content, char count \\
\texttt{decide\_start}    & Phase (VOLUNTARY or DECISION), prompt sent, calendar render \\
\texttt{decide\_end}      & Tool calls, thinking, usage, latency \\
\texttt{batch\_rejected}  & Attempt number, conflict description, rejected actions \\
\texttt{batch\_applied}   & Applied actions, calendar render after \\
\bottomrule
\end{tabular}
\end{table*}

\noindent The \texttt{system\_prompt} field of each \texttt{agent\_registered}
event contains the exact text delivered to the model, including any
adversarial or DSPy variant appended section.  Replaying a trace therefore
requires only the model API; all prompts are self-contained in the file.

\section{Model configurations}
\label{app:models}

Table~\ref{tab:model-configs} lists the seven models evaluated in the benchmark-lite experiments, along with their API provider and model identifier as used in the experiment configuration files.

\begin{table*}[t]
\centering
\small
\begin{tabular}{llll}
\toprule
Model name & Provider & API / access & Model identifier \\
\midrule
Claude Sonnet 4.6   & Anthropic & Google Vertex AI (Anthropic)   & \texttt{claude-sonnet-4-6} \\
DeepSeek V4 Pro     & DeepSeek  & OpenRouter                     & \texttt{deepseek/deepseek-v4-pro} \\
Gemini 3 Flash      & Google    & Google Vertex AI               & \texttt{gemini-3-flash-preview} \\
Gemini 3.1 Pro      & Google    & Google Vertex AI               & \texttt{gemini-3.1-pro-preview} \\
GPT-5.4 Mini        & OpenAI    & OpenAI API                     & \texttt{gpt-5.4-mini} \\
Llama 4 Maverick    & Meta      & Google Vertex AI (OpenAI-compat.) & \texttt{llama-4-maverick-17b-128e-instruct-maas} \\
Qwen3.6 Plus        & Alibaba   & OpenRouter                     & \texttt{qwen/qwen3.6-plus} \\
\bottomrule
\end{tabular}
\caption{Model identifiers and API providers for all seven models in the benchmark-lite evaluation. All models are accessed via the LiteLLM routing layer; the API format column reflects the provider-specific adapter used by the CalBench harness.}
\label{tab:model-configs}
\end{table*}

All cooperative benchmark-lite runs reported in the main table use the same prompt scaffold and game configuration (5 agents, 3 participants per meeting, 5 meetings per task, 16 time slots, 15 turns per round, no DM cap), with homogeneous same-model teams and private agent-to-agent DMs. The current harness also supports heterogeneous model teams by assigning a model provider and identifier per agent, as well as meeting-participant groupchat and all-agent groupchat communication conditions. The adversarial condition uses agent 0 with a frozen DSPy-compiled probing prompt (variant \texttt{c006}) against the same-model agents on the 5-agent, 3-participant subset (6 turns per round, DM cap 100).

\section{API Costs}

\begin{table}[h]
\centering
\caption{Estimated API costs for the canonical mixed-agent trace set. The four shared models appear in all 270 traces; Llama, Qwen, and DeepSeek appear in one 90-trace slice each. Costs assume non-cached, non-batch API pricing and bill output as completion plus reasoning tokens where reasoning-token usage is reported.}
\label{tab:api-cost-estimates}
\footnotesize
\setlength{\tabcolsep}{3.5pt}
\begin{tabular*}{\columnwidth}{@{\extracolsep{\fill}}lrrrr@{}}
\toprule
Model & Traces & Uniform & Varied & Total \\
\midrule
Gemini 3.1 Pro & 270 & $\$42.61$ & $\$41.81$ & $\$84.42$ \\
Claude Sonnet 4.6 & 270 & $\sim\$97.45$ & $\sim\$94.44$ & $\sim\$191.89$ \\
Gemini 3 Flash & 270 & $\$13.98$ & $\$13.81$ & $\$27.80$ \\
GPT-5.4 Mini & 270 & $\$17.67$ & $\$17.40$ & $\$35.07$ \\
Llama 4 Maverick & 90 & $\$2.73$ & $\$2.64$ & $\$5.37$ \\
Qwen3.6 Plus & 90 & $\$7.07$ & $\$6.95$ & $\$14.02$ \\
DeepSeek V4 Pro & 90 & $\$4.09$ & $\$3.98$ & $\$8.07$ \\
\bottomrule
\end{tabular*}
\end{table}

Claude traces did not record provider token counts, so Claude costs are estimated from reconstructed prompt and response lengths. Two Gemini 3 Flash calls with missing token usage are estimated the same way; all other rows use recorded token counts.

\section{Non-LLM Reference Protocols}
\label{app:baselines}
 
This appendix specifies the four non-LLM reference protocols introduced in \S\ref{sec:baselines}: their per-round message flows, slot-selection rules, parameters, and privacy-relevant invariants. All four share the CalBench harness (per-round cheap-talk, voluntary, decision, and resolution phases; same DM and broadcast channels; same \textsc{schedule}/\textsc{reschedule} action validation against local state). Differences below are entirely in the policy that maps observations to messages and actions.

\begin{table*}[t]
\centering
\small
\begin{tabular}{lp{0.72\textwidth}}
\toprule
Protocol & Rationale for inclusion \\
\midrule
\textsc{IMAP} & High-disclosure, low-regret reference point. Participants reveal full local cost vectors, so the initiator can choose the minimum-cost feasible slot for the current meeting. \\
\textsc{SD-MAP} & Low-disclosure feasibility reference point adapted from multi-agent meeting scheduling. Agents exchange binary proposal status rather than costs, exposing what is lost when the protocol avoids utility disclosure. \\
\textsc{DSM-Welfare} & Mechanism-design reference point with broad offers and welfare-leaning parameters, representing the high-coordination side of the DSM privacy--cost frontier. \\
\textsc{DSM-Private} & Matched DSM variant with narrow offers and a high privacy penalty, isolating the effect of pushing the same mechanism toward low disclosure. \\
\bottomrule
\end{tabular}
\caption{Why each non-LLM reference protocol is included. The protocols are not intended to model natural-language dialogue; they provide interpretable operating points for comparing LLM agents under the same private-information scheduling task.}
\label{tab:reference-protocol-rationale}
\end{table*}
 
\subsection{Incremental MAP (IMAP)}
\label{app:imap}
 
\paragraph{Protocol.} For each meeting, the lowest-id participant acts as initiator. The initiator sends a single \textsc{cost\_request} DM to every other participant, listing the slot indices of the calendar. Each responder computes its local insertion cost for every slot --- $0$ if the slot is free, the errand displacement cost if a movable errand can be displaced to a known free slot, and infeasible otherwise --- and returns the full vector in a single \textsc{costs} DM. The initiator sums per-slot costs across itself and all responders, picks the slot with minimum total feasible cost (breaking ties by slot index), and broadcasts a \textsc{decision} DM. All participants then schedule the agreed slot in the decision phase.
 
\paragraph{Cost-optimality and limits.} Because every participant reveals its full per-slot cost vector to the initiator, \textsc{IMAP} achieves the minimum-cost feasible insertion for the \emph{current} meeting whenever one exists in the local-errand subproblem. It treats previously scheduled multi-agent meetings as hard commitments and does not bump them: renegotiating a confirmed multi-agent meeting requires restarting a coordinated rescheduling episode \citep{modi2004multiagent}, which \textsc{IMAP} does not initiate. \textsc{IMAP} is therefore exact only with respect to the local errand-displacement subproblem of the current meeting.
 
\paragraph{Numeric disclosure footprint.} Two DMs per responder per meeting (one \textsc{cost\_request}, one \textsc{costs} reply). Each \textsc{costs} message carries one integer per slot index. There is no conditional or selective disclosure: the entire local cost structure is sent on every meeting.
 
\subsection{Scheduling-Difficulty MAP (SD-MAP)}
\label{app:sdmap}

\paragraph{Protocol.} \textsc{SD-MAP} is a low-disclosure reference protocol adapted from \citet{modi2004multiagent}'s scheduling-difficulty rescheduling heuristic. Meetings are processed in the incoming order defined by the task fixture; the protocol does not globally reorder meetings by difficulty. For each meeting, the lowest-id participant acts as initiator. The initiator iterates over candidate slots in local calendar order and sends a \textsc{propose} DM containing only the meeting id and proposed slot to every responder. Each responder evaluates the slot with a binary feasibility check: \textsc{PENDING} if the slot is free or holds a movable errand; \textsc{PENDING} if the slot holds a confirmed prior meeting that is bumpable under the scheduling-difficulty rule (in which case a tentative bump is recorded locally); \textsc{IMPOSSIBLE} otherwise. There are no cost vectors or score vectors --- feasibility is strictly binary. Replies arrive in a \textsc{reply} DM. If all replies are \textsc{PENDING}, the initiator broadcasts a \textsc{confirm} DM and the slot is agreed. Otherwise the initiator tries the next candidate slot. If no slot succeeds, the initiator sends a \textsc{fail} DM and the meeting is unresolved.

\paragraph{Bump repair via cheap-talk.} After a new meeting is confirmed, any tentative bump triggers a cheap-talk repair episode coordinated through the cheap-talk phase of the next round. The bumping participant sends a \textsc{reschedule\_request} DM to the bumped meeting's initiator. That initiator proposes a repair slot with a \textsc{propose\_reschedule} DM to all affected participants, who reply with \textsc{reschedule\_reply} (\textsc{PENDING} or \textsc{IMPOSSIBLE}). Once all affected participants accept, the initiator sends a \textsc{confirm\_reschedule} DM. Current participants then apply the agreed moves in their \textsc{decision} phase actions; non-participants apply them in the \textsc{voluntary} phase. If no repair slot can be found, the initiator sends \textsc{fail\_reschedule} and the bumped meeting falls back to its pre-bump slot.

\paragraph{The bumping rule.} \textsc{SD-MAP} has no slot score vector. Feasibility is binary: \textsc{PENDING} or \textsc{IMPOSSIBLE}. Scheduling difficulty $\Delta$ is a meeting-priority quantity used only for bump decisions, not slot ordering:
\[
\Delta(M) \;=\; \sum_{a \in \mathrm{participants}(M) \setminus \{\mathrm{self}\}} \sigma(a),
\]
where $\sigma : \mathcal{A} \to \mathbb{R}_{\geq 0}$ is a per-agent difficulty weight. A confirmed prior meeting $M_2$ is tentatively bumped in favor of the new meeting $M_1$ only when
\[
\textsc{Bump}(M_2 \mid M_1) \;\equiv\; \Delta(M_2) < \Delta(M_1),
\]
i.e., rescheduling the existing meeting is strictly easier than rescheduling the new one.

\paragraph{SD model.} CalBench supplies $\sigma$ via a configurable \texttt{sd\_model} entry on the game config; absent any specification all agents default to unit difficulty, which reduces the rule to ``bump iff the existing meeting has strictly fewer other participants than the new meeting.'' We use this default for all reported \textsc{SD-MAP} results.

\paragraph{Numeric disclosure footprint.} One \textsc{propose} DM per responder per attempted slot, and one binary \textsc{reply} in return. Successful confirmation adds one \textsc{confirm} DM per responder. Bump repair adds one \textsc{reschedule\_request}, one \textsc{propose\_reschedule} per affected participant, one \textsc{reschedule\_reply} per participant, and one \textsc{confirm\_reschedule} or \textsc{fail\_reschedule}. No costs, cost vectors, or slot indices beyond the one being proposed are ever transmitted.
 
\subsection{Distributed Score-based Multi-round (DSM)}
\label{app:dsm}
 
\paragraph{Origin.} The two \textsc{DSM} reference protocols (\textsc{DSM-Welfare} and \textsc{DSM-Private}) are implemented using the \textsc{PaperDSM} client, which follows the Distributed Score-based Multi-round mechanism of \citet{farhadi2021faithful} including satisfaction-level scoring, score-cost bookkeeping, and the flexibility-adjusted reward to the selected responder. We extend the protocol with multiple proposal sub-rounds per meeting: when the initial offer set yields no jointly feasible slot, the initiator proposes the next untried batch, repeating until a feasible slot is found or the slot pool is exhausted.
 
\paragraph{Protocol.} Every meeting has a designated initiator (the lowest-id participant). In each sub-round, the initiator selects an offer set of $L$ candidate slots from those it can locally schedule into, ordered by its own local feasibility score, and sends them as a single \textsc{proposals} DM to every responder. Each responder maps each proposed slot to a discrete satisfaction level $s \in \{0, 1, \dots, D{-}1\}$ with $D = 12$, where $0$ encodes infeasibility, $D{-}1$ encodes a free slot with no displacement, and intermediate levels decrease monotonically with displacement cost. Responders return the full score vector in a \textsc{scores} DM. The initiator combines local and reported scores into a per-slot rank and either announces the best fully-feasible slot via a \textsc{decision} DM, or proposes another untried batch. The protocol also supports cross-meeting displacement: when a candidate slot is blocked by a prior multi-agent meeting, the initiator can attach a proposed displacement plan to its offer, which the bumped meeting's participants score as additional responders \citep{modi2004multiagent}. All participants then commit the agreed slot and any agreed reschedules in the decision phase.
 
\paragraph{Satisfaction levels and scoring cost.} Each non-zero score $s$ carries a scoring cost $C(s) = D - s - 1$ that the responder pays to the initiator's point budget at decision time. The initiator pays back a flexibility-adjusted reward to the responder whose score was selected; responders whose offered slot is not chosen receive no reward but still pay $C(s)$ on each non-zero score they reported, mirroring the paper's faithfulness construction.
 
\paragraph{Offer-size utility.} The initiator chooses its offer size $L$ at each sub-round to maximize expected utility:
\[
\begin{aligned}
U(L) \;=\;&\;
p_{\text{succ}}(L)\,\bar v(L)
\;+\;
w_{\text{soc}}\, p_{\text{succ}}(L)
\\[0.2em]
&\;
-\;
\theta\, c_{\text{priv}}\, L
\;-\;
\beta \bigl(1 - p_{\text{succ}}(L)\bigr)
\end{aligned}
\]
over $L \in [L_{\min}, L_{\max}]$, where:
\begin{itemize}
  \item $p_{\text{succ}}(L) = 1 - (1 - \hat p)^L$ is the estimated probability that at least one of the top-$L$ candidate slots is jointly feasible. $\hat p$ is a density-based estimate of single-responder feasibility, obtained from the initiator's own calendar density and the responder count.
  \item $\bar v(L)$ is the mean local feasibility score of the top-$L$ candidates under the initiator's local cost ordering, normalized to $[0, 1]$.
  \item $\theta \geq 0$ scales a per-offer privacy cost (each disclosed slot is treated as an additional unit of leakage).
  \item $\beta \geq 0$ scales the cost of an extra negotiation round when the current offer fails to yield a fully feasible slot.
  \item $w_{\text{soc}} \geq 0$ is a social-welfare bonus on success.
\end{itemize}
The $\beta/\theta$ ratio is the dominant operating-point control: large $\beta$ relative to $\theta$ favors broad offer sets (\textsc{DSM-Welfare}), while large $\theta$ relative to $\beta$ favors narrow offer sets (\textsc{DSM-Private}).
 
\paragraph{Aggregate scoring statistics.} For a score vector $(s_1, \dots, s_L)$ returned by a responder, we compute three summary statistics used by the reward and observability machinery: \emph{availability} $A = |\{i : s_i > 0\}|$, the number of feasible offers; \emph{flexibility}, a base-$(A{+}1)$ polynomial encoding of the score multiset that monotonically rewards both more feasible offers and higher satisfaction within them; and \emph{scoring cost} $\sum_i C(s_i)$, the total points the responder pays. The initiator's selection reward to the responder whose score was chosen is then
\[
R \;=\; (D - 1 - s^\star) \;+\; \max\bigl(0, \min(A, L) - 1\bigr),
\]
where $s^\star$ is the selected score; the second term is the flexibility bonus that ensures responders with multiple feasible offers are not penalized for cooperativeness. We clip $R = 0$ when $s^\star = 0$ (infeasible) or $s^\star = D-1$ (free slot, no concession to compensate). The initiator's point budget is updated as $\Delta = \sum C(s_i) - R$ across all responders for that decision, preserving the faithful-mechanism property that points flow from those who concede less to those who concede more.
 
\paragraph{Numeric disclosure footprint.} One \textsc{proposals} DM per responder per sub-round (carrying $L$ slot indices) and one \textsc{scores} reply (carrying $L$ values in $\{0, \dots, D{-}1\}$). The number of sub-rounds per meeting is bounded by the search policy: \textsc{DSM-Welfare} runs exhaustively until a fully-feasible slot is found or the slot pool is exhausted; \textsc{DSM-Private} stops early under its tighter $\theta$. Cross-meeting displacement attempts add one \textsc{scores} DM per displaced-meeting participant per offer set in which the displacement appears.
 
\paragraph{Preset values.} Table~\ref{tab:dsm-presets} summarizes the parameter settings for the two DSM presets. \textsc{DSM-Welfare} uses CalBench's defaults; \textsc{DSM-Private} overrides them to anchor the high-privacy end of the frontier.
 
\begin{table*}[t]
\centering
\small
\begin{tabular}{lcc}
\toprule
Parameter & \textsc{DSM-Welfare} & \textsc{DSM-Private} \\
\midrule
Client type & \texttt{paper\_dsm} & \texttt{private\_dsm} \\
$L_{\min}$ & 1 & 1 \\
$L_{\max}$ (\texttt{dsm\_num\_proposals}) & 12 & 2 \\
$\beta$ (failure penalty) & 1.0 & 0.25 \\
$\theta$ (per-offer privacy cost) & 0.0 & 10.0 \\
$w_{\text{soc}}$ (social-welfare weight) & 1.0 & 0.25 \\
Cascade depth & 2 & 1 \\
Displacement targets considered & 4 & 2 \\
Exhaustive search of feasible offers & yes & no \\
\bottomrule
\end{tabular}
\caption{DSM preset parameters for the benchmark-lite reference-protocol runs. \textsc{DSM-Welfare} uses the \texttt{paper\_dsm} client with social-welfare-leaning settings from \citet{farhadi2021faithful}. \textsc{DSM-Private} uses the \texttt{private\_dsm} client, which internally overrides $L_{\max}$, $\beta$, $\theta$, $w_{\text{soc}}$, cascade depth, and displacement targets to minimize per-meeting disclosure.}
\label{tab:dsm-presets}
\end{table*}
 
\subsection{Common privacy invariants}
 
All four non-LLM reference protocols share two structural privacy properties relative to LLM agents. First, the message vocabulary is restricted to typed JSON: cost vectors (\textsc{IMAP}), proposal/reply/confirm/reschedule status (\textsc{SD-MAP}), or quantized scores over candidate slots (DSM). None transmit any natural-language content. Second, by construction none can leak the natural-language semantic context attached to calendar entries (low/medium/high sensitivity, \S\ref{sec:env}), since semantic context is never present in the message vocabulary. The protocols therefore set a quantitative floor of zero on context leakage, against which we measure the leakage rate of LLM agents communicating in unrestricted English over the same task.


\section{OpenSkill Leaderboard Ranking}
\label{app:openskill-ranking}

The hosted leaderboard uses OpenSkill \citep{Joshy2024}, a multiplayer rating
system related to TrueSkill \citep{herbrich2006trueskill}. Each calendar trace
is treated as a free-for-all rating event over the agent seats in that game.
Seats are mapped back to stable player identities by model name, so a model can
receive rating updates from many heterogeneous teams. For each model \(i\), we
maintain a separate rating distribution for each leaderboard metric,
\((\mu_{i,k},\sigma_{i,k})\), where \(\mu\) is estimated skill and \(\sigma\)
is uncertainty.

We rate models on three dimensions:
\[
k \in \{\mathrm{coordination}, \mathrm{cost}, \mathrm{privacy}\}.
\]
Coordination is higher-is-better; excess cost and excess VPS are
lower-is-better. Each dimension is updated separately, so the leaderboard does
not collapse a game to a single win/loss outcome: a model can gain coordination
rating while losing privacy rating in the same trace.

\paragraph{Score-margin updates.}
The leaderboard uses OpenSkill's score-margin variant rather than a rank-only
update. Instead of discarding metric magnitudes after ranking, we pass the raw
per-seat scores to OpenSkill. Coordination uses the scheduled-meeting ratio
directly, while excess cost and excess VPS are negated so that larger OpenSkill
scores are always better. OpenSkill then applies a logarithmic margin factor to
larger score gaps, so a decisive cost or privacy difference can produce a
larger rating update than a narrow one.

The margin parameters are \(0.2\) for coordination, \(10.0\) for excess cost,
and \(5.0\) for excess VPS. These margins are calibrated to the observed
per-game score ranges rather than treated as tie tolerances. Varied-cost errand
values remain on their raw \(\{1,10,100\}\) scale; we do not remap them to
\(\{1,2,3\}\) for the rating update. Model privacy scores use
reflection-calibrated excess VPS; protocol-semantic VPS is used for non-LLM
reference-protocol reports.

\paragraph{Displayed values.}
The raw parenthetical values shown in the leaderboard are not OpenSkill rating
scores. They report interpretable means such as coordination ratio, raw
game-level excess cost, and excess VPS. The separate \(\mu \pm \sigma\)
columns expose the underlying metric-specific OpenSkill ratings, while MMR
provides a single number for matchmaking and headline ordering.

Non-LLM reference protocols are shown separately from model ratings and ranked
by raw metrics only: coordination descending, then excess cost ascending, then
excess VPS ascending. They are not mixed into the model OpenSkill pool because
they serve as protocol reference points rather than hosted matchmaking
competitors.

\section{VPS metric: implementation details}
\label{app:vps}

This appendix specifies the VPS estimators introduced in \S\ref{sec:vps}: protocol-semantic VPS for typed DSM/IMAP/SD-MAP reference-protocol messages and reflection-calibrated VPS for LLM agents. Both are measurement procedures rather than online mechanisms; they do not alter the agents' scheduling behavior.

\subsection{Belief representation and update rule}

For each $(round, target, observer)$ triple $(r, i, j)$, the metric maintains a belief vector $\mathrm{Bel}^{r}_{j \to i} \in [0, 1]^{|S|}$ initialized to the uniform prior $\mathrm{Bel}^{r,\text{prior}}_{j \to i}[k] = p_0 = 0.5$. Each visible DM $m$ in trace order is replayed and may produce one or more belief-update events, each a tuple $(k, e, \alpha, \texttt{source})$ specifying the slot $k$ the message implicates, the evidence value $e \in [0, 1]$ about feasibility (1.0 = ``slot is feasible'', 0.0 = ``slot is infeasible''), the strength $\alpha \in [0, 1]$ of the inference, and the source tag for downstream decomposition. The update is the strength-weighted nudge
\[
\mathrm{Bel}'[k] \;=\; (1 - \alpha)\, \mathrm{Bel}[k] \;+\; \alpha\, e,
\]
applied independently to each slot $k$. With $\alpha = 1$ the posterior at $k$ is replaced by $e$; with $\alpha = 0$ the posterior is unchanged; intermediate values produce a partial move. Updates are accumulated over the round, and end-of-round leakage is computed from Eq.~\ref{eq:vps}. We track each update's source so that VPS can be decomposed by message channel in the results.

\subsection{Typed-message update rules for reference protocols}

All non-LLM reference protocols emit typed JSON DMs with known semantics. Table~\ref{tab:vps-typed} gives the full protocol-semantic rule set. The visibility model is point-to-point: each DM contributes only to the belief pair defined by its sender and recipient. Most typed updates are exact; SD-MAP proposals receive a softer strength because proposing a slot is strong evidence of local availability but not logically identical to a hard feasibility report.

\begin{table*}[t]
\centering
\small
\begin{tabular}{llll}
\toprule
Source protocol & Message & Slot(s) updated & Evidence $e$ \\
\midrule
DSM   & \textsc{proposals} & each proposed slot                & 1.0 \\
DSM   & \textsc{scores}    & each scored slot                  & $0$ if score $\leq 0$, else $s/(D{-}1)$ \\
DSM   & \textsc{decision}  & chosen slot                       & 1.0 \\
IMAP  & \textsc{cost\_request} & (no update)                  & --- \\
IMAP  & \textsc{costs}     & each requested slot               & $0$ if cost is \texttt{None}, else 1.0 \\
IMAP  & \textsc{decision}  & chosen slot                       & 1.0 \\
SD    & \textsc{propose}             & proposed slot    & 0.85 \\
SD    & \textsc{reply}               & proposed slot    & $0$ if \textsc{IMPOSSIBLE}, else 1.0 \\
SD    & \textsc{propose\_reschedule} & target slot      & 0.85 \\
SD    & \textsc{reschedule\_reply}   & target slot      & $0$ if \textsc{IMPOSSIBLE}, else 1.0 \\
\bottomrule
\end{tabular}
\caption{Typed-message belief updates. DSM and IMAP rules use $\alpha = 1$. SD \textsc{propose} and \textsc{propose\_reschedule} use $\alpha = 0.70$, because proposing a slot is strong but not logically identical to a hard feasibility report; SD reply rules use $\alpha = 1$. \textsc{cost\_request} carries no evidence about the sender's calendar (the initiator is asking, not telling), so it produces no update; the corresponding \textsc{costs} reply is matched against the request to bound the slot set being scored. DSM \textsc{scores} replies are matched against the originating \textsc{proposals} message to recover slot indices from plan ids.}
\label{tab:vps-typed}
\end{table*}

DSM \textsc{scores} replies require care because the reply payload carries plan ids rather than slot indices; the parser maintains a $(round, sender, recipient) \to \texttt{plan\_id} \mapsto \texttt{slot}$ map populated by the corresponding \textsc{proposals} message and uses it to attribute scores to the right slots. IMAP \textsc{costs} replies are similarly bounded by the slot set in the originating \textsc{cost\_request}, so a responder that returns extra slots is not credited with extra leakage. Errand-occupied slots that admit a movable-displacement insertion are treated as feasible ($e = 1$) by IMAP \textsc{costs}, since the responder reports a finite cost rather than \texttt{None}; this matches IMAP's protocol semantics, where slot $k$ being feasible-via-displacement is exactly the information the message reveals.

\subsection{Reflection-calibrated VPS (LLM agents)}

LLM-agent cheap-talk is unstructured, so the main LLM privacy metric does not
attempt to recover beliefs from regex rules over messages. Instead, CalBench
records measurement-only reflection calls. After each round in the reported
mixed-team runs, each agent is asked, for every other target agent and every
slot, how its belief changed about whether the target is occupied. The prompt
requires a JSON array of signed integers in \(\{-3,-2,-1,0,1,2,3\}\), where
positive values mean ``more willing to say occupied'' and negative values mean
``less willing to say occupied.'' These calls use \texttt{record\_history=false}
and are not inserted into later game context.

The calibration pipeline first extracts the emitted deltas and, when available,
token log-probabilities for the numeric slot decisions. The VPS conversion is
slot-equivalent: a maximal \(\pm 3\) movement counts as one slot of raw belief
movement. We also compute calibrated loss by checking the direction of movement
against the target's actual calendar state: movement toward the truth counts as
privacy leakage, while wrong-direction movement counts as zero calibrated
leakage. Per-game target summaries subtract a fixed five-slot communication
floor and report \texttt{excess\_calibrated\_vps\_loss\_total}, which is the
privacy value used for model-based OpenSkill analyses.

\subsection{Slot-equivalent units and the prior}

\paragraph{Slot-equivalent units.} VPS treats every slot equally regardless of displacement cost. This matches the original VPS formulation \citep{maheswaran2006vps} and makes comparisons across uniform- and varied-cost calendars directly interpretable: a one-slot belief movement has the same privacy cost whether the occupied item is cheap or expensive to move.

\paragraph{Prior choice.} We use $p_0 = 0.5$ throughout, reflecting maximal observer uncertainty about an unfamiliar calendar at round start. Sensitivity to this choice is bounded: under any prior $p_0 \in [0.3, 0.7]$, the relative ordering of clients on VPS is unchanged in our data; results in the main text use $p_0 = 0.5$.

\subsection{Output schema and reporting}

The protocol-semantic pipeline emits four CSV tables per analysis run.

\paragraph{\texttt{belief\_evidence.csv}} Per-update rows. Columns: \texttt{trace\_path}, \texttt{game\_id}, \texttt{event\_index}, \texttt{round}, \texttt{target\_agent}, \texttt{observer\_agent}, \texttt{slot}, \texttt{source}, \texttt{evidence}, \texttt{strength}, \texttt{belief\_before}, \texttt{belief\_after}. This table supports source-level decomposition of exact typed-protocol leakage.

\paragraph{\texttt{pair\_round\_vps.csv}} Per-$(round, target, observer)$ rows. Columns: \texttt{trace\_path}, \texttt{game\_id}, \texttt{round}, \texttt{target\_agent}, \texttt{observer\_agent}, \texttt{target\_is\_participant}, \texttt{observer\_is\_participant}, \texttt{num\_agents}, \texttt{num\_slots}, \texttt{observations}, \texttt{prior\_distance\_to\_ideal}, \texttt{posterior\_distance\_to\_ideal}, \texttt{vps\_loss}, \texttt{vps\_loss\_per\_slot}, and \texttt{prior}.

\paragraph{\texttt{game\_summary.csv}} Per-$(trace, game)$ rows. Aggregates pair-round rows: \texttt{vps\_loss\_total}, \texttt{vps\_loss\_mean}, \texttt{participant\_pair\_vps\_loss\_total}, \texttt{participant\_pair\_vps\_loss\_mean}, and \texttt{observation\_count}.

\paragraph{\texttt{game\_target\_summary.csv}} Per-$(trace, game, target)$ rows. This table is used for per-agent rating inputs and reports \texttt{vps\_loss\_total}, \texttt{excess\_vps\_loss\_total}, participant-pair totals, and the fixed floor subtracted from each game-target total.

\paragraph{Headline metrics.} In the main results we report mean per-game VPS loss as the primary privacy-leakage axis for direct comparability across uniform- and varied-cost calendars. We retain participant-pair-only summaries to distinguish within-meeting disclosure from accidental exposure involving non-participants.

\section{Additional Results}
\label{app:additional-results}

\begin{figure}[t]
  \centering
  \includegraphics[width=\linewidth]{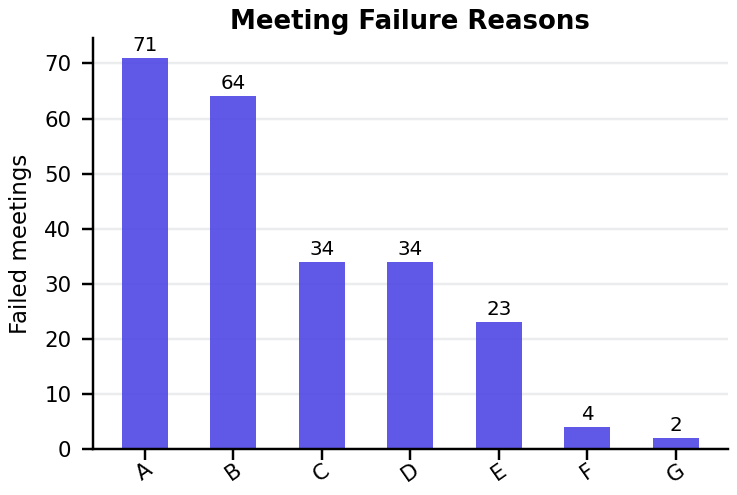}
  \caption{
    Meeting-level scheduling failure modes in the mixed-model CalBench traces.
    Each failed meeting is assigned one reporting reason.
    Letter labels denote:
    \textbf{A} = new meeting schedule mismatch;
    \textbf{B} = attempted blocked item reschedule;
    \textbf{C} = missing schedule action;
    \textbf{D} = meeting reschedule mismatch;
    \textbf{E} = new meeting schedule slot occupied;
    \textbf{F} = reschedule slot occupied;
    \textbf{G} = multiple reasons.
  }
\label{fig:appendix-failure-modes}
\end{figure}

\subsection{Completion-conditioned regret}
\label{app:completion-conditioned-regret}

Table~\ref{tab:completion-conditioned-regret} reports excess cost among
agent-game seats that completed all of their assigned meetings. This isolates
whether agents schedule well after conditioning on feasibility. Across the
canonical mixed-agent cohort, 358 of 630 evaluated agent-game seats completed
every assigned meeting. Among those full-completion seats, 26.3\% still incurred
nonzero excess cost relative to the oracle, and the mean excess cost was 1.20
per scheduled meeting.

\begin{table}[t]
\centering
\footnotesize
\begin{adjustbox}{max width=\columnwidth}
\begin{tabular}{lrrr}
\toprule
Model & Full seats & Mean excess & Nonzero regret \\
\midrule
Claude Sonnet 4.6 & 52 & 0.76 & 11.5\% \\
DeepSeek V4 Pro & 62 & 0.27 & 19.4\% \\
Gemini 3 Flash & 52 & 0.31 & 26.9\% \\
Gemini 3.1 Pro & 50 & 0.85 & 28.0\% \\
GPT-5.4 Mini & 38 & 6.00 & 42.1\% \\
Llama 4 Maverick & 43 & 1.62 & 41.9\% \\
Qwen3.6 Plus & 61 & 0.27 & 23.0\% \\
\bottomrule
\end{tabular}
\end{adjustbox}
\caption{Completion-conditioned regret. Rows include only agent-game seats with
100\% completion. Mean excess is scheduled excess cost per successful meeting
within those full-completion seats. Nonzero regret is the share of
full-completion seats with any oracle excess cost.}
\label{tab:completion-conditioned-regret}
\end{table}

\subsection{First-speaker effects}
\label{app:first-speaker-effects}

The sequential meeting protocol creates an exposure asymmetry: the first
speaker often anchors the negotiation by proposing feasible slots or ruling out
bad ones before seeing how much other participants are willing to reveal. We
therefore stratify each participating agent-meeting by whether that agent was
first in the meeting's speaker order. For this diagnostic, the unit is a
participating agent-meeting: VPS is the round-level reflection-calibrated
excess VPS for that target agent, and excess cost is the scheduled
agent-meeting cost above the oracle schedule.

The aggregate effect is concentrated in privacy rather than welfare. Across all
models, first speakers incur 4.71 excess VPS compared with 3.78 for subsequent
speakers, a paired increase of 0.93 VPS per participating agent-meeting
(\(p=9.26\times10^{-24}\), Benjamini--Hochberg \(q=2.22\times10^{-22}\)).
The same comparison is significant in both uniform-cost games
(\(\Delta=1.03\), \(q=9.29\times10^{-14}\)) and varied-cost games
(\(\Delta=0.82\), \(q=1.33\times10^{-9}\)). In contrast, excess cost is
statistically indistinguishable by speaker role: descriptive means are 1.84 for
first speakers and 1.86 for subsequent speakers, and the paired aggregate test
is not significant (\(\Delta=0.10\), \(p=0.69\), \(q=0.83\)).

\begin{figure}[t]
\centering
\includegraphics[width=\linewidth]{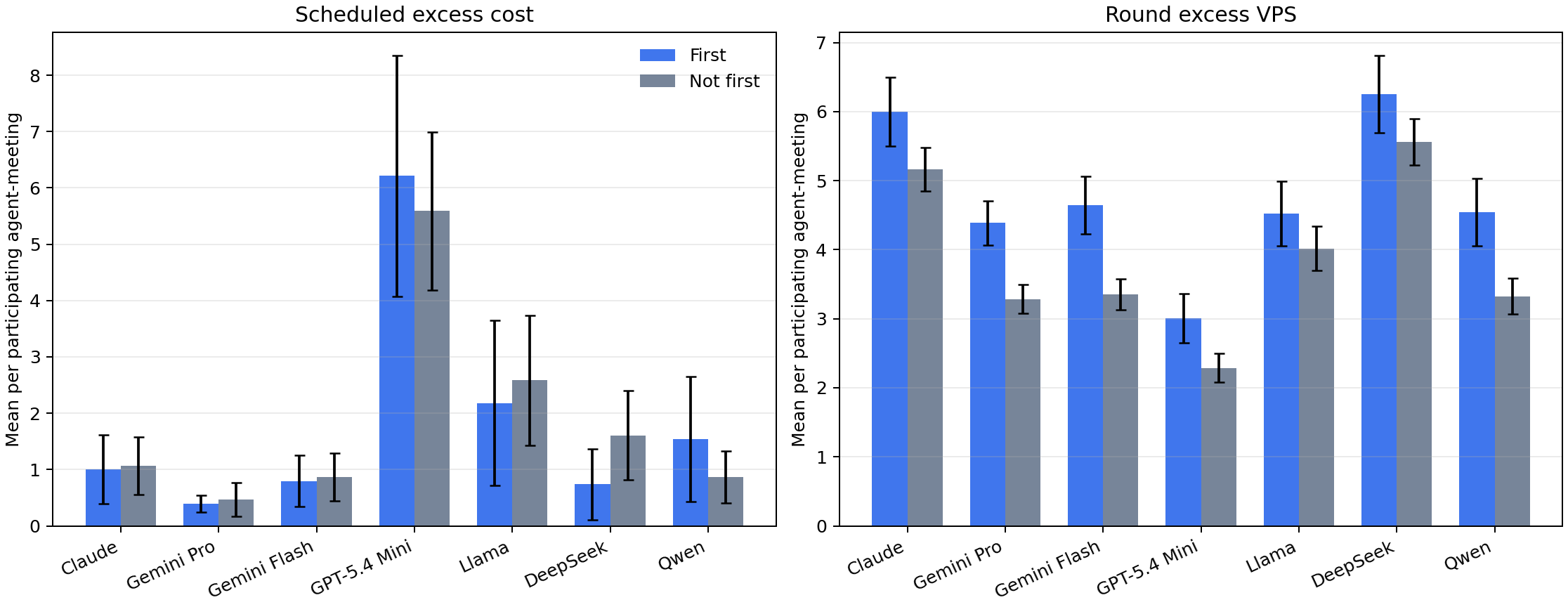}
\caption{First-speaker effects by model. Bars report mean scheduled excess cost
and mean round-level excess VPS per participating agent-meeting, stratified by
whether the agent was the first speaker for that meeting. Error bars are
95\% confidence intervals over participating agent-meeting rows.}
\label{fig:first-speaker-effects-by-model}
\end{figure}

\begin{table}[t]
\centering
\small
\begin{tabular}{lrrrr}
\toprule
Model & \(\Delta\) VPS & VPS \(q\) & \(\Delta\) cost & Cost \(q\) \\
\midrule
Claude & 0.83 & \textbf{0.006} & -0.03 & 0.952 \\
Gemini Pro & 1.10 & \textbf{\(2.84\times10^{-8}\)} & -0.10 & 0.832 \\
Gemini Flash & 1.30 & \textbf{\(5.13\times10^{-8}\)} & 0.07 & 0.901 \\
GPT-5.4 Mini & 0.72 & \textbf{\(2.32\times10^{-4}\)} & 1.31 & 0.727 \\
Llama & 0.50 & 0.058 & -0.46 & 0.832 \\
DeepSeek & 0.69 & \textbf{0.033} & -0.99 & 0.546 \\
Qwen & 1.22 & \textbf{\(1.44\times10^{-5}\)} & 0.67 & 0.727 \\
\bottomrule
\end{tabular}
\caption{Paired first-speaker tests by model. Each test pairs an agent-game's
mean first-speaker meetings with its mean non-first-speaker meetings. Values are
first minus subsequent speaker; \(q\)-values are Benjamini--Hochberg corrected
within metric.}
\label{tab:first-speaker-effects-by-model}
\end{table}

\subsection{Communication volume by meeting and first-speaker role}
\label{app:communication-volume}

\begin{figure}[t]
\centering
\includegraphics[width=\linewidth]{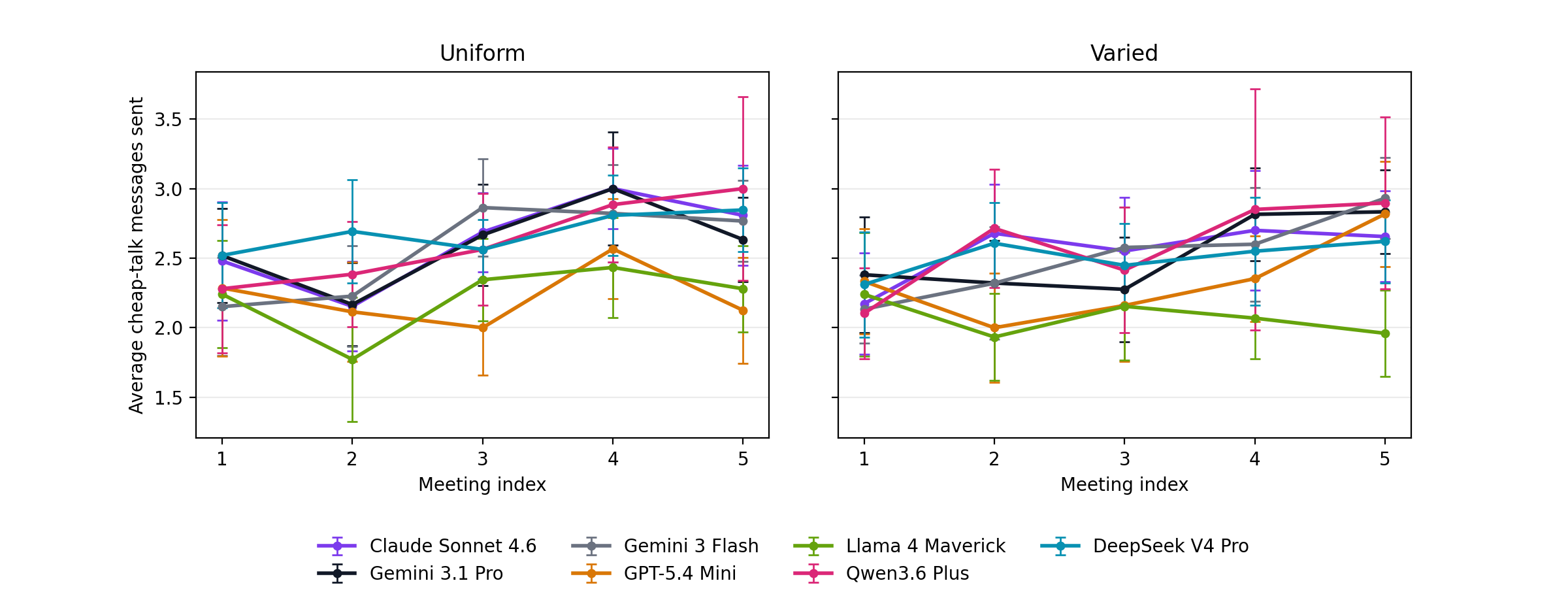}
\caption{Average cheap-talk messages sent by selected model seats at each meeting index in the canonical mixed-agent cohort. The metric is included because increasing message counts across meetings indicate whether agents accumulate unresolved coordination burden over the sequence rather than treating each meeting independently. Mixed5 traces contribute all five model seats; Qwen and DeepSeek arena-entry traces contribute entrant agent 0.}
\label{fig:avg-dms-by-meeting}
\end{figure}

\begin{figure}[t]
\centering
\includegraphics[width=\linewidth]{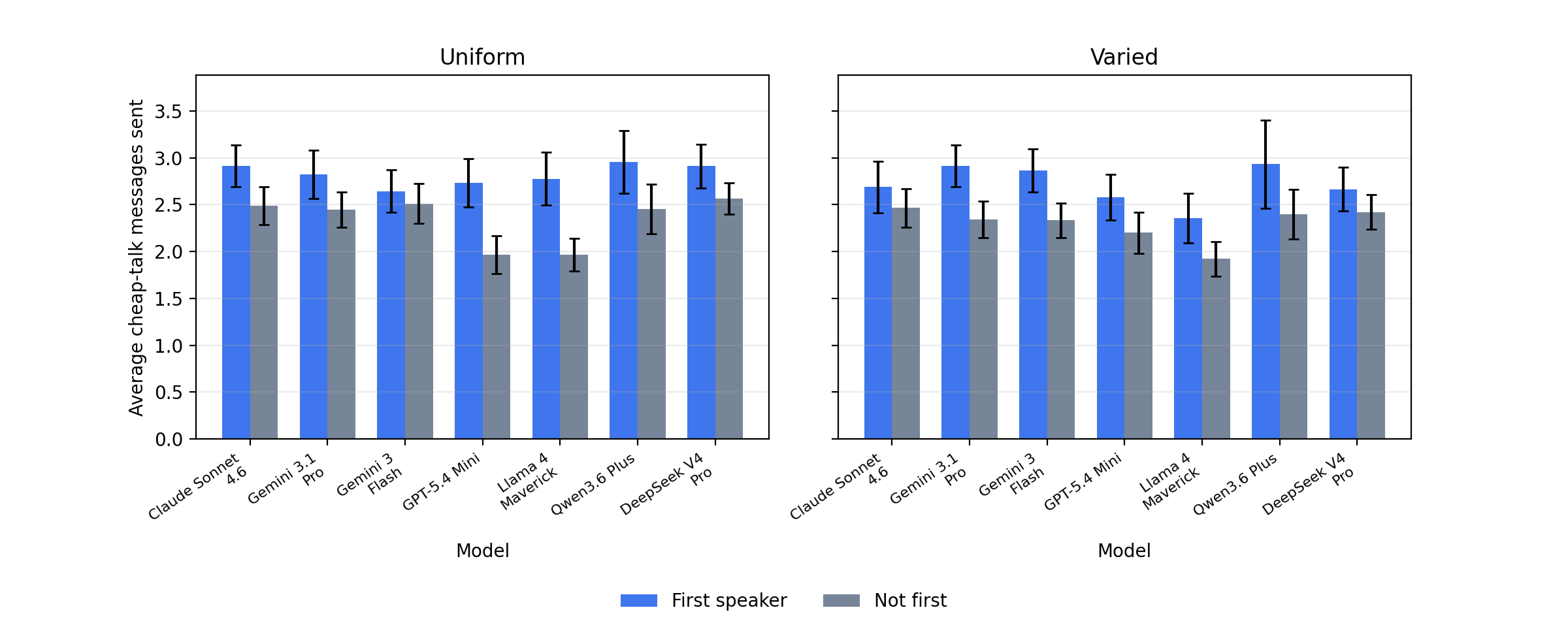}
\caption{Average cheap-talk messages sent by selected model seats, stratified by whether the seat was the first speaker for that meeting. We collapse second and third speaker positions because first-speaker exposure is balanced by task construction, whereas later speaker positions are confounded with fixed seat identity in the canonical task suite.}
\label{fig:avg-dms-by-first-speaker-role}
\end{figure}

\subsection{Prior-meeting rescheduling}
\label{app:prior-meetings}

\begin{figure}[t]
  \centering
  \includegraphics[width=\linewidth]{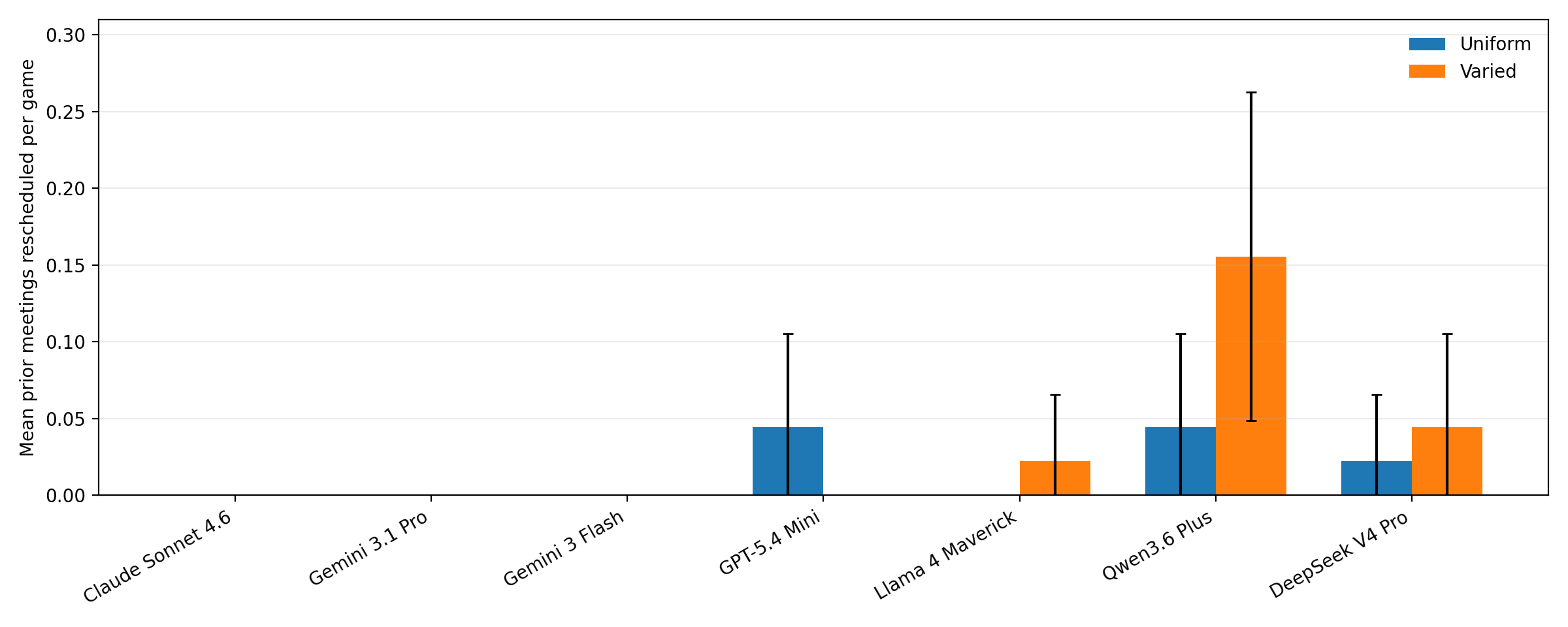}
  \caption{
    Mean number of prior meetings rescheduled per game by selected model seats in the canonical mixed-agent cohort. Higher values indicate more frequent cross-meeting repair, where an agent moves an already scheduled multi-agent meeting to accommodate a later one.
  }
  \label{fig:prior-meeting-reschedule-rate}
\end{figure}

Figure~\ref{fig:prior-meeting-reschedule-rate} reports how often agents move previously scheduled multi-agent meetings while coordinating later meetings. For each mixed-cohort game, we count the number of distinct prior meetings rescheduled by the selected model seat, then average this count across games for each model and setting. This metric captures rare but important cross-meeting repair behavior: rather than resolving conflicts only by moving local errands or selecting a different free slot, agents sometimes reopen an earlier multi-agent commitment to reduce disruption.

\subsection{Validation of own-slot disclosure regexes}
\label{app:own-slot-regex-validation}

The fairness analysis in \S\ref{sec:results} uses a deterministic message audit
to measure whether agents propose their own slots and whether they disclose
cost- or constraint-relevant information about those slots. This audit is
separate from the reflection-calibrated VPS metric: it is a targeted diagnostic
used to interpret GPT-5.4 Mini's fairness behavior. The extractor first detects
slot mentions, then requires first-person language (e.g., ``I'', ``my
calendar'', ``for me'') before assigning one of three labels: own-slot proposal,
own-slot availability, or own cost/constraint disclosure. The cost/constraint
label includes qualitative cost language such as ``difficult'', ``cheap'',
``blocked'', ``can move'', and ``move my errand''; exact numeric cost mentions
are tracked separately and are rare.

To estimate extraction quality, we manually annotated a balanced validation
sample drawn from the varied-cost mixed-model traces. For each label, we sampled
15 regex-positive messages and 15 regex-negative messages, then annotated whether
the message actually contained the target phenomenon under the definitions
above. Table~\ref{tab:own-slot-regex-validation} reports precision and recall
on this audit sample with Wilson 95\% confidence intervals. The regexes are
intentionally conservative: precision is high for own-slot proposal and
availability, but recall is moderate because the rules miss elliptical
statements such as ``both 12 and 13 are feasible for me'' or ``slot 4 it is;
I'll make the necessary adjustment.'' The cost/constraint detector has lower
precision because some messages mention another agent's cost or constraint in a
sentence that also contains first-person language. These errors make the main
claim conservative with respect to GPT-5.4 Mini: GPT-5.4 Mini is not lower on
own-slot proposals, and its much lower cost/constraint disclosure remains large
despite recall limits.

\begin{table}[t]
\centering
\small
\begin{tabular}{lcc}
\toprule
Regex label & Precision & Recall \\
\midrule
Own-slot proposal & 1.00 [0.80, 1.00] & 0.65 [0.45, 0.81] \\
Own-slot availability & 1.00 [0.80, 1.00] & 0.56 [0.37, 0.72] \\
Own cost/constraint & 0.80 [0.55, 0.93] & 0.63 [0.41, 0.81] \\
\bottomrule
\end{tabular}
\caption{Manual validation of the regex labels used in the own-slot disclosure
audit. Bracketed intervals are Wilson 95\% confidence intervals computed on a
balanced audit sample of 15 regex-positive and 15 regex-negative messages per
label.}
\label{tab:own-slot-regex-validation}
\end{table}

\section{Homogeneous-team Behavioral Trace Audit}
\label{app:homogeneous-audit}
\label{app:emergent-patterns}

The behavioral taxonomy reported here comes from an earlier homogeneous-agent
audit suite, including the probe condition used to surface semantic-context
leakage. These traces are useful for identifying recurring language and
coordination failures, but they are not used for the headline VPS comparisons in
the main text because they do not include reflection-calibrated VPS scores.
Table~\ref{tab:homogeneous-audit-provenance} summarizes the role of this suite.

\begin{table}[t]
\centering
\small
\begin{tabular}{lp{0.56\linewidth}}
\toprule
Property & Value \\
\midrule
Team composition & Homogeneous model teams \\
Games & 504 \\
Messages & 36{,}858 DMs \\
Models & 7 model families \\
Available signals & Message text, actions, trace-state detectors \\
Unavailable signals & Reflection-calibrated VPS \\
Use in paper & Behavioral taxonomy and mechanism audit \\
\bottomrule
\end{tabular}
\caption{Provenance for the homogeneous-team trace audit. Counts from this suite
are reported separately from the mixed-agent benchmark results.}
\label{tab:homogeneous-audit-provenance}
\end{table}

Table~\ref{tab:emergent-full} reports all twelve emergent patterns, ranked by
occurrence count. Detection is non-exclusive: a single message may match multiple
patterns.

\emph{Errand mention} (M, rank~3): messages referencing ``errand'' as a calendar item type, revealing structure beyond abstract availability. Concentrated in Gemini~3~Flash (2{,}058), Llama (1{,}548), and Qwen (862); nearly absent in Claude (34) and GPT (0).

\emph{Cost gossip} (S, rank~4): an agent relays another agent's displacement costs or errand details to a third party---a compounded privacy violation. Gemini~3~Flash (678) and Gemini~3.1~Pro (516) are the primary sources.

\emph{Cost-number disclosure} (M, rank~5): messages containing explicit numerical cost values (e.g., ``cost of 100''), a direct system-prompt violation. Gemini~3.1~Pro accounts for 89\% of all instances (712 of 797).

\emph{Unilateral meeting displacement} (S, rank~8): an agent moves a previously scheduled meeting without coordinating with its other participants, causing a consistency violation. The dominant coordination failure mode; Qwen~3.6~Plus is most affected (33 of 74 instances).

\emph{Errand-ID leakage} (M, rank~7): messages containing internal identifiers such as ``Errand~\#10.'' Almost entirely a Llama issue (56 of 133 instances in varied-cost traces).

\emph{Semantic label leakage} (M, rank~10): messages revealing personal event descriptions (e.g., ``medical appointment,'' ``library drop-off''). The rarest (35 total) but most sensitive privacy violation.

\emph{Cascading failure} (S, rank~11): a single consistency violation propagates across multiple rounds. In one Qwen trace, a Meeting~M2 displacement caused three consecutive round failures.

\emph{Output truncation} (S, rank~12): an agent's JSON response is cut off before the actions array, producing a null scheduling decision. Almost exclusively Llama (10 of 12 instances), caused by ~65-token response truncation.

\begin{table*}[t]
\centering
\small
\begin{adjustbox}{max width=\textwidth}
\begin{tabular}{clrllp{4.5cm}}
\toprule
Rank & Pattern & Count & Level & Top models & Detection method \\
\midrule
1 & Availability gossip & 6{,}971 & Social & DeepSeek (1{,}782), Claude (1{,}643), Flash (1{,}067) & Third-agent ID + availability keyword within 60 chars \\
2 & Confirmation tax & 6{,}537 & Msg & Flash (1{,}428), Gem.~Pro (1{,}317), GPT (939) & $<80$ chars + agreement phrase match \\
3 & Errand mention & 5{,}981 & Msg & Flash (2{,}058), Llama (1{,}548), Qwen (862) & Substring match: ``errand'' \\
4 & Cost gossip & 1{,}757 & Social & Flash (678), Gem.~Pro (516), Llama (245) & Third-agent ID + cost/errand keyword within 60 chars \\
5 & Cost-number disclosure & 797 & Msg & Gem.~Pro (712), Flash (72), Llama (12) & Regex: \texttt{cost[= ]+\textbackslash d+} and variants \\
6 & Qualitative flattening & 311 & Msg & GPT (102), Llama (71), DeepSeek (44) & Cost $\geq 100$ errand displaced in varied-cost trace \\
7 & Errand-ID leakage & 133 & Msg & Llama (56), others scattered & Regex: \texttt{[Ee]rrand\textbackslash s*\#\textbackslash d+} \\
8 & Unilateral displacement & 74 & Social & Qwen (33), Llama (18), GPT (15) & Non-empty consistency-violation field \\
9 & Ambiguous consensus & 37 & Social & GPT (21), Llama (9), DeepSeek (7) & $\geq 2$ distinct non-null slots in failed resolution \\
10 & Semantic label leakage & 35 & Msg & DeepSeek (7), Qwen (2), Flash (1) & 27 curated semantic-context keywords \\
11 & Cascading failure & 21 & Social & Qwen, GPT, Llama & Same meeting ID violated in $>1$ resolution per trace \\
12 & Output truncation & 12 & Social & Llama (10), DeepSeek (2) & Agent slot is \texttt{null} in failed resolution \\
\bottomrule
\end{tabular}
\end{adjustbox}
\caption{Full taxonomy of emergent behavioral patterns across 504 homogeneous-team games (36{,}858 DMs, 7 models), ranked by occurrence. \emph{Level}: Message (individual message content) or Social (multi-agent dynamics). Detection is non-exclusive.}
\label{tab:emergent-full}
\end{table*}

\section{Post-hoc LLM-as-judge privacy audit}
\label{app:llm-judge-privacy-audit}

This appendix records the external LLM-as-judge audit we ran as a robustness
check on the main VPS measurements. The audit is not the primary privacy metric:
protocol-semantic VPS remains the primary measurement for typed non-LLM
reference protocols, and reflection-calibrated VPS remains the primary
measurement for LLM agents. Instead, the audit asks whether an independent
language model, given only the messages delivered to a recipient, detects a
similar pattern of calendar-state leakage.

\subsection{Corpus and artifacts}

We audited 180 initial-calibration games: the 90-game DSM-private baseline run
and the 90-game mixed-agent model run used for the main cohort analysis. We did
not include the later live-matchmaking games. The audited trace roots were:
\begin{itemize}[leftmargin=*]
    \item \url{calbench-mixed/baseline-dsm-private-calbench-mixed-001}
    \item \url{calbench-mixed/mixed5-calbench90-001}
\end{itemize}

We also preserved the run outputs in
S3 under
\url{s3://calbench-traces/calbench-mixed/_analysis/posthoc-judge-vps-180-20260525/}.
The most important generated tables are:
\begin{itemize}[leftmargin=*]
    \item \texttt{judge\_beliefs.csv}: row-level judge belief deltas.
    \item \texttt{judge\_cases.json}: the delivered-message cases shown to the judge.
    \item \texttt{judge\_beliefs\_calibrated.csv}: judge deltas with truth-direction calibration.
    \item \texttt{metric\_comparison.csv}: aggregate comparison against protocol-semantic and reflection-calibrated VPS.
    \item \texttt{judge\_vs\_reflection\_precision.csv}: per-game-target comparison of judge and reflection precision.
    \item \texttt{judge\_more\_precise\_than\_reflection.csv}: filtered cases where the judge's nonzero deltas were more often truth-directed than self-reflection.
    \item \texttt{judge\_reflection\_cell\_agreement.csv}: slot-belief cell agreement between judge and self-reflection.
\end{itemize}

\subsection{Judge task}

The judge script was
\url{analysis/scripts/round_message_judge_vps.py}. We used the Vertex AI
model \url{publishers/google/models/gemini-3.1-pro-preview}. Each judge API
call corresponded to one recipient's delivered transcript for one game round,
not to one slot. The prompt included the messages that were visible to that
recipient in that round, including point-to-point DMs, participant group-chat
messages, and all-agent broadcast messages. The judge then emitted belief
updates for the other agents' 16 calendar slots.

The output scale matched the self-reflection VPS scale:
\[
\Delta \in \{-3,-2,-1,0,1,2,3\},
\]
where positive values mean the recipient should become more willing to believe
that the target agent is occupied in that slot, negative values mean less
willing to believe occupied, and zero means no belief movement. Raw judge VPS
uses the slot-equivalent conversion \( |\Delta|/3 \). Calibrated judge VPS counts
only movements in the truth-directed direction, using the target agent's actual
calendar state.

Overall, the audit created 2{,}702 judge cases and 9{,}721 nonzero judged belief
rows. The total raw judge VPS was 8{,}999.33 slot-equivalents, or 3.33
slot-equivalents per judged case on average. The mean absolute nonzero judged
belief delta was 2.78 on the seven-point scale.

\subsection{Aggregate comparison}

Table~\ref{tab:llm-judge-vps-aggregate} summarizes the aggregate audit results.
For DSM-private, the comparison point is the protocol-semantic VPS parser for
typed DSM messages. For mixed5, the comparison point is the existing
reflection-calibrated VPS measurement collected during the game runs.

\begin{table*}[t]
\centering
\small
\begin{tabular}{lrrrr}
\toprule
Dataset & Beliefs & Raw judge VPS & Calibrated judge VPS & Primary metric VPS \\
\midrule
DSM-private & 1{,}851 & 1{,}344.67 & 327.67 & 863.64 \\
Mixed5 & 7{,}870 & 7{,}654.67 & 4{,}490.33 & 1{,}752.17 \\
\bottomrule
\end{tabular}
\caption{External judge VPS compared with the primary privacy measurement for each audited dataset. The DSM-private primary metric is protocol-semantic uniform VPS. The mixed5 primary metric is reflection-calibrated VPS.}
\label{tab:llm-judge-vps-aggregate}
\end{table*}

The mixed5 judge found substantially more calibrated leakage than
self-reflection: 4{,}490.33 calibrated slot-equivalents versus 1{,}752.17 from
reflection-calibrated VPS. After subtracting the five-slot per-game-target
communication floor, the mixed5 judge excess calibrated VPS was 2{,}552.33,
compared with 1{,}491.33 for self-reflection. This suggests that the external
judge is more sensitive to message-level implications than the acting models'
own reported belief updates.

DSM-private shows the opposite calibration pattern: the judge's raw VPS is
higher than the protocol-semantic parser, but its calibrated VPS is much lower.
This is expected to be less directly interpretable because DSM messages expose
typed feasibility and score information, while the judge is asked to express
belief changes about occupied slots in natural language terms. Protocol-semantic
parsing is therefore the appropriate primary measure for DSM.

\subsection{Truth-direction precision}

For nonzero judge deltas, 4{,}639 of 7{,}870 mixed5 belief rows moved in the
truth-directed direction, giving a nonzero precision of 0.589. The corresponding
self-reflection precision was 0.528 over 10{,}445 nonzero self-reflection slot
deltas. At the game-target level, 447 targets could be joined between the two
measurement procedures. The external judge was more precise than self-reflection
for 256 game-targets, self-reflection was more precise for 182 game-targets, and
the two were tied for 9 game-targets. The mean precision gap, judge minus
self-reflection, was 0.016 overall and 0.174 among targets where the judge was
more precise.

These results are useful for interpreting the role of self-reflection. The
external judge is often a sharper detector of explicit message implications, but
self-reflection still captures the model's own reported belief movement. The two
measurements are therefore complementary rather than redundant.

\subsection{Cell-level agreement with self-reflection}

We also computed agreement between the external judge and self-reflection at the
slot-belief cell level, using the key
\((\mathrm{game\_id}, \mathrm{observer\_agent}, \mathrm{target\_agent},
\mathrm{slot})\). Repeated rows for the same cell were summed before comparison,
with mean deltas also retained in the output table. Table~\ref{tab:llm-judge-reflection-agreement}
reports the main agreement statistics.

\begin{table}[h]
\centering
\small
\begin{tabular}{lr}
\toprule
Agreement statistic & Value \\
\midrule
Union cells compared & 24{,}727 \\
Judge nonzero cells & 6{,}828 \\
Self-reflection nonzero cells & 7{,}914 \\
Both nonzero & 4{,}472 \\
Nonzero-support Jaccard & 0.435 \\
Judge coverage of reflection nonzero & 0.565 \\
Reflection coverage of judge nonzero & 0.655 \\
Directional agreement when both nonzero & 0.643 \\
Cohen's \(\kappa\), signs over full union & 0.353 \\
Pearson signed-delta correlation, full union & 0.290 \\
Pearson signed-delta correlation, both nonzero & 0.426 \\
\bottomrule
\end{tabular}
\caption{Slot-belief cell agreement between post-hoc judge deltas and self-reflection deltas for the mixed5 90-game run.}
\label{tab:llm-judge-reflection-agreement}
\end{table}

Agreement is weak-to-moderate rather than interchangeable. The two methods
share a meaningful signal: when both mark a slot-belief cell as nonzero, they
agree on direction 64.3\% of the time, and the signed-delta correlation rises to
0.426. However, the support overlap is incomplete, with 2{,}356 judge-only
nonzero cells and 3{,}442 reflection-only nonzero cells. This supports the
interpretation that self-reflection is not merely a noisy external annotation
proxy. It measures an internally reported belief update, while the external
judge provides an independent audit of what an outside reader could infer from
the same delivered messages.

\subsection{Takeaways}

The audit supports three conclusions. First, external LLM judging can be run on
both typed protocol agents and LLM agents because it only requires delivered
messages and trace-state calibration, although typed protocol agents are still
best measured by their protocol semantics. Second, for LLM cheap-talk, the judge
detects more calibrated leakage than self-reflection, especially in explicit
availability and free-slot disclosures. Third, the moderate but incomplete
agreement between judge and self-reflection justifies using reflection-calibrated
VPS as the main model-based privacy metric while reporting external judging as a
robustness audit rather than a replacement.


\end{document}